\documentclass{article}%
\usepackage{amssymb}%
\usepackage{amsmath}%
\usepackage{amsfonts}%
\usepackage{graphicx}
\providecommand{\U}[1]{\protect\rule{.1in}{.1in}}
\begin{document}

\title{On Extracting Physical Content from Asymptotically Flat Space-Time Metrics}
\author{C. Kozameh$^{1}$, E. T. Newman$^{2}$, \ G. Silva-Ortigoza$^{3}$\\$^{1}$FaMaF, Univ. of Cordoba, \\Cordoba, Argentina\\$^{2}$Dept of Physics and Astronomy, \\Univ. of Pittsburgh, \\Pittsburgh, PA 15260, USA\\$^{3}$Facultad de Ciencias F\'{\i}sico Matem\'{a}ticas \\de la Universidad Aut\'{o}noma de Puebla, \\Apartado Postal 1152, 72001,\\Puebla, Pue., M\'{e}xico}
\date{2.12.08}
\maketitle

\begin{abstract}
A major issue in general relativity, from its earliest days to the present, is
how to extract physical information from any solution or class of solutions to
the Einstein equations. Though certain information can be obtained for
arbitrary solutions, e.g., via geodesic deviation, in general, because of the
coordinate freedom, it is often hard or impossible to do. Most of the time
information is found from special conditions, e.g., degenerate principle null
vectors, weak fields close to Minkowski space (using coordinates close to
Minkowski coordinates) or from solutions that have symmetries or approximate
symmetries. In the present work we will be concerned with asymptotically flat
space times where the approximate symmetry is the Bondi-Metzner-Sachs (BMS)
group. For these spaces the Bondi four-momentum vector and its evolution,
found from the Weyl tensor at infinity, describes the total energy-momentum of
the interior source and the energy-momentum radiated. By generalizing the
structures (shear-free null geodesic congruences) associated with the
algebraically special metrics to asymptotically shear-free null geodesic
congruences, which are available in all asymptotically flat space-times, we
give kinematic meaning to the Bondi four-momentum. In other words we describe
the Bondi vector and its evolution in terms of a center of mass position
vector, its velocity and a spin-vector, all having clear geometric meaning.
Among other items, from dynamic arguments, we define a unique (at $our$ level
of approximation) total angular momentum and extract its evolution equation in
the form of a conservation law with an angular momentum flux.

\end{abstract}

\section{ Introduction}

From the very earliest days of general relativity (GR) the issue of extracting
the physical meaning or content in the solutions of the Einstein equations has
been a difficult problem\cite{Weyl} and usually has been solved only in
special cases or special situations, e.g., in the presence of symmetries or
approximate (asymptotic) symmetries. Probably the best studied case is that of
asymptotically flat space-times where an asymptotic symmetry exists, namely
the BMS group, and where, at null infinity, the total energy-momentum of the
interior source was identified as well as gravitational radiation i.e., energy
and momentum loss. \ These identifications, in terms of the asymptotic
components of the Weyl tensor, were made via group theoretical arguments
combined with physical insight into the dynamics of the radiation process.
However the problem of gravitational multipole moments,\ and in particular,
the mass dipole and the angular momentum has proved to be difficult. \ At the
present time, using symmetry arguments, there are several different proposed
definitions\cite{Szabados,Helfer,dray,Winicour}of these quantities with little
apparent use of dynamic considerations. \ (To our knowledge only in linear
theory or in stationary cases has the issue of quadrupole moments been
addressed.)\ \ In the present work, using largely a \textit{dynamic argument}
and verifying it by a symmetry consideration, we return to the issue of
recognizing the mass-dipole moment and angular momentum hidden in the
asymptotic Weyl tensor. From this dynamic argument the mass-dipole moment and
angular momentum, using the Bondi energy-momentum four-vector, are found. Both
kinematic expressions, as well dynamic equations (equations of motion), are determined.

In flat-space Maxwell theory the related issues are much simpler though some
questions do remain. \ The charge, obtained by a 2-D surface integral of the
asymptotic Maxwell field is a constant and lies in the scalar representation
of the Lorentz group; the dipole moments (electric and magnetic) also obtained
by 2-D surface integrals of the asymptotic Maxwell field (though the results
depend on a time direction and the 2-surface chosen for the integral) also lie
in a finite dimensional representation of the Lorentz group. \ For a static or
stationary system it is easy to define the center of charge (so that the
electric dipole moment associated with it vanishes) and a constant magnetic
dipole moment. \ Recently it was shown\cite{shearfreemax} that for general
asymptotically flat Maxwell fields (with non-vanishing total charge) one could
find a unique complex world-line in complex Minkowski space, referred to as
the complex center of charge, from which the dynamic electric and magnetic
moments could be obtained.

It is this construction of the complex center of charge for the Maxwell field
that we generalize to GR. This leads, in a very unorthodox manner, to a
complex center of mass. The real part is identified with the usual center of
mass while the imaginary part is the specific spin-angular momentum, i.e., the
spin per unit mass. The argument leading to this complex world line is based
largely on dynamics and analogies with electrodynamics and special solutions
of GR, e.g., the algebraically special metrics and in particular, the charged
spinning metric\cite{kerr-newman}. It is applied to both the asymptotic vacuum
Einstein equations and the Einstein-Maxwell equations. For ease of
presentation most of the analysis will be for the vacuum case. The results for
the Einstein-Maxwell case will be presented towards the end of Sec. III
without detailed derivation.

In order to get a better perspective on this work and on our point of view,
several comments might be of some use.

Both Maxwell theory and GR are considered to be fundamental physical theories
[in addition to quantum theory which so far appears incompatible with GR] and
thus in principle this construction is applicable to the gravitational field
in the neighborhood of any massive body, from elementary particles to
galaxies. Though GR is almost always applied to astrophysical situations,
there is in principle no reason that it could not be applied at some level to
laboratory masses or even to elementary particles. \ If the individual masses
are sufficiently far apart we could even consider their gravitational fields
at (relatively) large distances as being asymptotically flat and use the
general theory of asymptotically flat space-times for their analysis. The
theory of asymptotically flat space-times appears to be the best tool to
define isolated bodies in GR and, as in Newton's theory, the notion of
isolated bodies is an approximation of reality that is useful in general, and
specifically for defining center of mass, intrinsic angular momentum, etc.

It is this point of view we adopt in this work. \ Though there is no rigorous
way to make physical identification with the GR variables we will consider
three very similar alternative identifications - some more intuitive than
others - and see what are the theoretical consequences - and then argue for
the last one. The idea is then to look at the dynamic predictions of each of
the alternatives. At the linear level they all agree: it is in the
consideration of the non-linear terms that we see the differences and see how
one of the alternatives, probably the least intuitive one, comes closest to
what we would physically expect.

The starting point for the discussion are the asymptotic Bianchi identities.
We have assumed that the Einstein or the Einstein/Maxwell equations have been
integrated in the asymptotic region using the standard
Bondi-Sachs-Newman-Penrose
asymptotic\cite{bondi,sachs,newman-penrose,newman-unti}(peeling theorem)
behavior. \ When the relevant integrations have been completed we are left
with several evolution equations, namely the asymptotic Bianchi Identities
(the Bondi supplementary conditions) and asymptotic Maxwell
equations\cite{newman-tod}, that become our object for analysis and interpretation.

In Sec. II we review the relevant ideas about asymptotically flat space-times
(Einstein or Einstein/Maxwell) at null infinity\cite{newman-tod} . More
specifically we discuss the structure of null infinity (and its
complexification), i.e., \ Penrose's $\mathfrak{I}^{+}.$ We then review a
variety of subjects: e.g., what `lives' on $\mathfrak{I}^{+},$ the asymptotic
Weyl tensor with the Bianchi Identities and the Bondi shear, other related
structures as cuts of $\mathfrak{I}^{+}$, past-light-cones and angle fields.
The spherical harmonic decomposition of the various fields is then described.
In this context we discuss the Bondi four-momentum and the energy/momentum
loss theorem.\qquad

In Sec. III, we examine three alternative suggestions for extracting physical
information from the Weyl tensor components in the asymptotic Bianchi
Identities. A central structure in this extraction process comes from the
dynamics associated with the Bondi four-momentum. More specifically we want
\textit{first} to identify the mass dipole (or center of mass) and angular
momentum. Then, second, we want to establish dynamical laws for these
quantities, i.e., we want equations of motion for the center of mass and
evolution equations - with fluxes - for the angular momentum. They come
directly from the energy-momentum loss equations and a certain reality
condition. These dynamical laws are given at future null infinity and
therefore are intended to apply to the entire \textit{isolated} system.
Interactions between different bodies can not be considered in this framework.

The first and most `obvious' or intuitive model for the selection for the mass
dipole and angular momentum turns out to lack certain essential dynamic
features and has little or no geometric meaning. We consider the first (and
second) approaches as `toy' models, introduced in order motivate and clarify
the third and primary model. The second and the third approaches which are
based on specific choices of complex slicing of future null infinity, referred
to as (complex) cut functions, lead to very similar dynamics. The nod goes to
the third choice, it being more `natural' in its similarity or analogy to the
algebraically special metrics and to its unity and geometric attractiveness.
They are both based on a geometric construction of complex cut functions. One
of them, the second, is more appropriate to flat space, the other (which does
reduce to the second in flat space) is far better suited to the general
asymptotically flat case. In the later case we give an extended description of
the predictions and dynamic results including the results for the
Einstein-Maxwell equations. These results include kinematic expressions for
the Bondi mass and momentum, a center of mass and its equations of motion and
a definition of angular momentum that includes both spin and orbital terms
with a conservation law with an angular momentum flux.

In Sec. IV we analyze the invariance, under the BMS group, of the third
approach to this method of assigning physical meaning to the variables. We
briefly discuss the representation theory of the Lorentz group due to Gelfand,
Graev and Vilinkin\cite{GGV,held-newman}, and in particular apply their
description of the invariant finite dimensional subspaces of the infinite
dimensional representations, to our physical assignments. As a simple example,
we apply the representation theory to the Bondi energy-momentum four-vector
and show that indeed it is a Lorentzian four-vector. The details are given in
appendix 7.2.

In the discussion section, Sec. V, we summarize and discuss our results,
including the results for the Einstein-Maxwell case. \ We also speculate on
possible consequences of these results. One of our more interesting results -
worth speculating on - is that a simple geometric condition on the
Einstein-Maxwell fields leads to the Dirac value of the gyromagnetic ratio.

Our method for the study of the physical identifications is based on the use
of spherical harmonic expansions so that the coefficients in these expansions
become identified with physical objects. It is clear from the non-linearity of
the theory that it is impossible to work with the exact infinite expansions
and, thus, we must work with a truncated series. Specifically we expand
everything up to and including the $l=2$ harmonics and then include only
2$^{nd}$ order products. (In one place, where the result is physically so
attractive, we have included a 3$^{rd}$ order term.) Almost all of the
non-linear terms in our results arise from the (frequent) use of
Clebsch-Gordon expansions of spherical harmonic products.

\section{Preliminaries}

\subsection{\noindent\noindent\noindent Null Infinity}

After integrating\cite{bondi,sachs,newman-penrose,newman-unti,newman-tod} the
Einstein (or Einstein-Maxwell) equations along null surfaces in the
spin-coefficient formalism for large affine parameter, $r$, we are left, in
the limit of large $r$ with several evolution equations\cite{newman-tod} (for
the asymptotic Weyl tensor components) that are referred to as the
\textquotedblleft asymptotic Bianchi identities\textquotedblright. \ (These
equations are analogous to the remaining Maxwell equations\cite{newman-tod}
after all the radial integrations have been performed.) This limit of large
$r$ has been formalized by Penrose into the idea of the future null boundary
of space-time, a null three-surface, and referred to as $\mathfrak{I}^{+}.$
Our remaining equations are thus a set of differential equations on
$\mathfrak{I}^{+}.$ This null boundary has the topology of $S^{2}\times R$ and
an assignment of (Bondi) coordinates ($u,\zeta,\overline{\zeta}$ ), with $u$
on the $R$ part and ($\zeta,\overline{\zeta})$ as complex stereographic
coordinates on the $S^{2}.$ The freedom in the choice of these Bondi
coordinates, known as the BMS group is geometrically the asymptotic symmetry
group. The BMS group is composed of two parts, the supertranslations%

\begin{align}
\widehat{u}  & =u+\alpha(\zeta,\overline{\zeta})\label{s.t.}\\
(\widehat{\zeta},\overline{\widehat{\zeta}})  & =(\zeta,\overline{\zeta
})\nonumber
\end{align}
with $\alpha(\zeta,\overline{\zeta})$ an arbitrary smooth function on the
sphere and the Lorentz
transformations\cite{bondi,sachs,sachs2,penrose1,held-newman} given by
\begin{align}
\widehat{u}  & =Ku\label{l.t.}\\
K  & =\frac{1+\zeta\overline{\zeta}}{(a\zeta+b)(\overline{a}\overline{ \zeta
}+\overline{b})+(c\zeta+d)(\overline{c}\overline{\zeta}+\overline{d}%
)}\nonumber\\
\widehat{\zeta}  & =\frac{a\zeta+b}{c\zeta+d};\text{ \ }ad-bc=1.\nonumber
\end{align}

Though we are dealing with real space-times and $\mathfrak{I}^{+}$ is a real
three-surface, since we assume that all the relevant functions are analytic,
it is useful to allow the $u$ to take on complex values close to the real and
to allow $\overline{\zeta}$ to deviate slightly from the complex conjugate of
$\zeta.$

In addition to the Bondi coordinates we also have a null tetrad $(l^{a}%
,n^{a},m^{a},\overline{m}^{a})$ that is associated with the Bondi coordinates
(and does change with a BMS coordinate transformation) where $n$ is tangent to
the null generators of $\mathfrak{I}^{+},$ $m^{a}$ and $\overline{m}^{a}$ are
tangent to the $u=$ constant slices' of $\mathfrak{I} ^{+}$ while $l$ is the
null vector normal to the slices pointing into the space-time along the null
surface $u$ $=$ constant.

\subsection{\noindent Further Structures}

There are several important structures associated with $\mathfrak{I}^{+}$ that
we now describe.

$\bullet$ In addition to the Bondi `slicing' of $\mathfrak{I}^{+}$ given by
the `cuts', $u=$ constant, one can take an arbitrary real one-parameter, $s$,
family of slices given by real cut functions,%

\begin{equation}
u=G(s,\zeta,\overline{\zeta}),\label{G}%
\end{equation}
or their generalization to analytic complex \textit{cut functions}
\begin{equation}
u=X(\tau,\zeta,\overline{\zeta})\label{X}%
\end{equation}
with inverse functions
\[
\tau=T(u,\zeta,\overline{\zeta})
\]
where the complex $\tau$ must be able to be chosen so that $u$ is real. The
freedom in the choice of the parameter $\tau$: $\widehat{\tau}=F(\tau)$
$\ $with $F$ analytic, will be used later to normalize a physical variable.

$\bullet$ At every point of $\mathfrak{I}^{+}$ there is the past light-cone of
rays going back into the interior. The directions are labeled by the complex
stereographic angle ($L,\overline{L}$) with the zero value taken along $l$ and
the infinity along $n.$ Any complex angle field $L(\zeta,\overline{\zeta}),$
i.e., a stereographic angle given for each point on $\mathfrak{I}^{+},$ can be
expressed\cite{footprints} in terms of a complex cut function, restricted to
real $u^{\prime}$s by
\begin{align}
L(u,\zeta,\overline{\zeta})  & =\eth _{(\tau)}X(\tau,\zeta,\overline{\zeta
})\label{L=ethX}\\
\tau & =T(u,\zeta,\overline{\zeta}).\label{T}%
\end{align}
The subscript ($\tau$) means the application of $\eth $\ holding $\tau$ constant.

$\bullet$ We will have considerable use for the local Lorentz transformation
(null rotations) at each point of $\mathfrak{I}^{+}$ parametrized by the
arbitrary angle field%

\[
L=L(u,\zeta,\overline{\zeta})
\]
that preserve the vector $n,$ namely%

\begin{align}
l^{*}  & =l+L\overline{m}+\overline{L}m+L\overline{L}n\label{null rot}\\
m^{*}  & =m+Ln\nonumber\\
n^{*}  & =n\nonumber
\end{align}
For each point on $\mathfrak{I}^{+},$ the null vector $l^{*}$ determines a
null geodesic extending backwards into the space-time so that the field of
$l^{*\prime}$s determines a null geodesic congruence of the space-time.

\subsection{\noindent What lives on $\mathfrak{I}^{+}?$}

The following functions are defined\cite{newman-tod} on $\mathfrak{I}^{+};$

$\bullet$ $\sigma=\sigma(u,\zeta,\overline{\zeta}),$ the asymptotic shear of
the null geodesic congruence with Bondi tangent vector $l.$

$\bullet$ $\psi_{1}^{0\,}(u,\zeta,\overline{\zeta})$ and $\psi_{2}%
^{0\,}(u,\zeta,\overline{\zeta})$ are the leading terms of two tetrad
components of the Weyl tensor, $\psi_{1}^{\,}=\psi_{1}^{0\,}/r^{4}$ +.... and
$\psi_{2}^{\,}=\psi_{2}^{0\,}/r^{3}+...$

with%

\begin{align*}
\psi_{1}^{\,}  & =-C_{abcd}l^{a}m^{b}l^{c}m^{d}\\
\psi_{2}^{\,}  & =-C_{abcd}\overline{m}^{a}n^{b}l^{c}m^{d}%
\end{align*}

Under the tetrad transformation, Eq.(\ref{null rot}), the Weyl components
$\psi_{1}^{0\,}$ and $\psi_{2}^{0\,}$ transform as
\begin{align}
\psi_{1}^{0\ast\,}  & =\psi_{1}^{0\,}-3L\psi_{2}^{0\,}+3L^{2}\text{$\eth $%
(}\overline{\sigma})^{\cdot}+L^{3}\overline{\sigma}^{\cdot\cdot}%
\label{psi trans 1}\\
\psi_{2}^{0\ast\,}  & =\psi_{2}^{0\,}-2L\text{$\eth $(}\overline{\sigma
})^{\cdot}-L^{2}\overline{\sigma}^{\cdot\cdot}\label{psi trans 2}%
\end{align}

The $\psi_{1}^{0\,}$ and $\psi_{2}^{0\,}$ satisfy the asymptotic Bianchi Identities%

\begin{align}
(\psi_{1}^{0\,})^{\cdot}  & =-\text{$\eth $}\psi_{2}^{0}+2\sigma
\text{$\eth $(}\overline{\sigma})^{\cdot}\label{1}\\
(\psi_{2}^{0\,})^{\cdot}  & =-\text{$\eth $}^{2}\text{(}\overline{\sigma
})^{\cdot}-\sigma(\overline{\sigma})^{\cdot\cdot}\label{2}\\
\psi_{2}^{0\,}-\overline{\psi}_{2}^{0\,}  & =\overline{\text{$\eth $}}%
^{2}\sigma-\text{$\eth $}^{2}\overline{\sigma}+(\sigma)^{\cdot}\overline
{\sigma}-\text{(}\overline{\sigma})^{\cdot}\sigma.\label{3}%
\end{align}

The last two equations can be rewritten in terms of the `\textit{mass aspect}
', $\Psi,$ as%

\begin{align}
\Psi^{\cdot}  & =\sigma^{\cdot}\overline{\sigma}^{\cdot},\label{2*}\\
\Psi & =\overline{\Psi}=\psi_{2}^{0\,}+\text{$\eth $}^{2}\overline{\sigma
}+\sigma(\overline{\sigma})^{\cdot}.\label{3*}%
\end{align}
When a Maxwell field is present, these equations become modified and the
Maxwell equations must be included\cite{newman-tod} :%

\begin{align}
(\psi_{1}^{0\,})^{\cdot}  & =-\text{$\eth $}\psi_{2}^{0}+2\sigma
\text{$\eth $(}\overline{\sigma})^{\cdot}+2k\phi_{1}^{0}\overline{\phi}%
_{2}^{0}\label{Bondi1}\\
(\psi_{2}^{0\,})^{\cdot}  & =-\text{$\eth $}^{2}\text{(}\overline{\sigma
})^{\cdot}-\sigma(\overline{\sigma})^{\cdot\cdot}+k\phi_{2}^{0}\overline{\phi
}_{2}^{0}\label{Bondi3}\\
\overline{\Psi}  & =\Psi=\psi_{2}^{0\,}+\text{$\eth $}^{2}\overline{\sigma
}+\sigma(\overline{\sigma})^{\cdot},\label{Bondi6}\\
k  & =2Gc^{-4}%
\end{align}%
\begin{align}
& (\phi_{0}^{0})^{\cdot}+\text{$\eth $}\phi_{1}^{0}-\sigma\phi_{2}%
^{0}=0,\label{phiB00cdot}\\
& (\phi_{1}^{0})^{\cdot}+\text{$\eth $}\phi_{2}^{0}=0,\label{phiB10cdot}%
\end{align}

\subsection{\noindent Asymptotically Shear-free Null Geodesic Congruences}

$\bullet$ If the angle field $L(u,\zeta,\overline{\zeta})$ satisfies the
differential equation\cite{aronson,footprints}
\begin{equation}
\text{$\eth $}L+LL^{\cdot}=\sigma(u,\zeta,\overline{\zeta})\label{shearfree}%
\end{equation}
the null geodesic congruence determined by the null vector field given by
Eq.(\ref{null rot}) is asymptotically shear-free. It has been shown
earlier\cite{footprints} that solutions to Eq.(\ref{shearfree}) that are
regular on $\mathfrak{I}^{+}$, i.e., have no infinities, are given by the
following construction:

$L(u,\zeta,\overline{\zeta})$ is given parametrically by
\begin{align}
L(u,\zeta,\overline{\zeta})  & =\text{$\eth $}_{(\tau)}X(\tau,\zeta
,\overline{\zeta})\label{reg.sol}\\
u  & =X(\tau,\zeta,\overline{\zeta})
\end{align}
where $X(\tau,\zeta,\overline{\zeta})$ is found by first solving the `good
cut' equation\cite{hspace},
\begin{equation}
\text{$\eth $}^{2}Z=\sigma(Z,\zeta,\overline{\zeta})\label{g.c.}%
\end{equation}
whose solutions are known to depend on four arbitrary complex parameters,
$z^{a},$ i.e., $Z=Z(z^{a},\zeta,\overline{\zeta}).$ By choosing an arbitrary
world-line in the parameter space, ($H$-space\cite{hspace}), i.e., $z^{a}%
=\xi^{a}(\tau),$ $X(\tau,\zeta,\overline{\zeta})$ is determined by%

\begin{equation}
u=X(\tau,\zeta,\overline{\zeta})=Z(\xi^{a}(\tau),\zeta,\overline{\zeta
}).\label{good cut}%
\end{equation}
We thus have that every regular solution to the asymptotically shear-free
condition, Eq.( \ref{shearfree}), is determined by an arbitrary complex
world-line in a four-complex dimensional parameter space. \ The freedom,
mentioned earlier, in the choice of $\tau$ is used later to give a
normalization to $v^{a}(\tau)\equiv\partial_{\tau}\xi^{a}.$

We could reverse the procedure just described and assume that $u=X(\tau
,\zeta,\overline{\zeta})$ was given rather than $\sigma(u,\zeta,\overline
{\zeta}).$ Then $\sigma(u,\zeta,\overline{\zeta})$ and $L(u,\zeta
,\overline{\zeta})$ could be determined \textit{parametrically} by
\begin{align}
u  & =X(\tau,\zeta,\overline{\zeta})\label{parametric1}\\
L(u,\zeta,\overline{\zeta})  & =\text{$\eth $}_{(\tau)}X(\tau,\zeta
,\overline{\zeta})\label{parametric2}\\
\sigma(u,\zeta,\overline{\zeta})  & =\text{$\eth $}_{(\tau)}^{2}X(\tau
,\zeta,\overline{\zeta})\label{parametric3}%
\end{align}
or more explicitly by the harmonic series,
\begin{align}
u  & =X(\tau,\zeta,\overline{\zeta})=\frac{1}{\sqrt{2}}\xi^{0}(\tau)-\frac
{1}{2}\xi^{i}(\tau)Y_{1i}^{0}(\zeta,\overline{\zeta})+\xi^{ij}(\tau
)Y_{2ij}^{0}(\zeta,\overline{\zeta})+...,\\
L(u,\zeta,\overline{\zeta})  & =\xi^{i}(\tau)Y_{1i}^{1}(\zeta,\overline{\zeta
})-6\xi^{ij}(\tau)Y_{2ij}^{1}(\zeta,\overline{\zeta})+...\\
\sigma(u,\zeta,\overline{\zeta})  & =24\xi^{ij}(\tau)Y_{2ij}^{2}+...,
\end{align}
with $\xi^{a}=(\xi^{0}(\tau),$ $\xi^{i}(\tau)).$The complex parameter $\tau$
is given by $\tau=T(u,\zeta,\overline{\zeta}),$ Eq.(\ref{T})$.$ Note the
important point that when $\tau$ is replaced by $T$, the spherical harmonic
decomposition of those variables becomes non-trivial since it involves
products of different spherical harmonics.

In the special case of flat-space, (with $\sigma(u,\zeta,\overline{\zeta}
)=0$), the \textit{asymptotically shear-free} congruences become
\textit{\ shear-free} congruences and the regular solutions to
Eq.(\ref{shearfree}), given parametrically, become%

\begin{align}
u  & =X(\tau,\zeta,\overline{\zeta})=\xi^{a}(\tau)\widehat{l}_{a}%
(\zeta,\overline{\zeta})=\frac{1}{\sqrt{2}}\xi^{0}(\tau)-\frac{1}{2}\xi
^{i}(\tau)Y_{1i}^{0}(\zeta,\overline{\zeta})\label{flat X}\\
L(u,\zeta,\overline{\zeta})  & =\text{$\eth $}X(\tau,\zeta,\overline{\zeta
})=\xi^{a}(\tau)m_{a}(\zeta,\overline{\zeta}),\label{flat L}\\
\widehat{l}_{a}(\zeta,\overline{\zeta})  & =\frac{\sqrt{2}}{2}(1,\frac
{\zeta+\overline{\zeta}}{1+\zeta\overline{\zeta}},-i\frac{\zeta-\overline
{\zeta}}{1+\zeta\overline{\zeta}},\frac{-1+\zeta\overline{\zeta}}%
{1+\zeta\overline{\zeta}}),\\
\widehat{m}_{a}(\zeta,\overline{\zeta})  & =\text{$\eth $}\widehat{l}%
_{a}(\zeta,\overline{\zeta})=\frac{\sqrt{2}}{2P}(0,1-\overline{\zeta}%
^{2},-i(1+\overline{\zeta}^{2}),\text{ }2\overline{\zeta}),
\end{align}
with $\widehat{l}_{a}(\zeta,\overline{\zeta})$ a flat-space null vector on the
`light-cone' composed of the $l=0,1$ harmonics. In other words the regular
flat-space shear-free null geodesic congruences are determined by an analytic
complex curve in complex Minkowski space. \ One can reinterpret the Minkowski
space curve as a curve in the space of asymptotic complex Poincare
translations. The advantage of this latter interpretation is that it applies
just as well to the asymptotically flat space-times.

\subsection{\noindent Spherical Harmonic Decomposition}

All the functions on $\mathfrak{I}^{+}$ that we are dealing with have a spin
weight ($s$) and most have a definite conformal weight ($w$). They thus can be
expanded in the spin-weighted tensor harmonics-\cite{s-harmonics}. Denoting
the spin and conformal weights by $(s,w)$ as a subscript e.g., $W_{(s,w)},$ we have%

\begin{align}
u  & =X\equiv X_{(0,1)}=\frac{1}{\sqrt{2}}\xi^{0}(\tau)-\frac{1}{2}\xi
^{i}(\tau)Y_{1i}^{0}(\zeta,\overline{\zeta})+\xi^{ij}(\tau)Y_{2ij}^{0}%
(\zeta,\overline{\zeta})+...\label{X*}\\
\sigma & \equiv\sigma_{(2,-2)}=24\xi^{ij}(\tau)Y_{2ij}^{2}+...\label{sigma*}\\
L  & \equiv L_{(1,\times)}=\xi^{i}(\tau)Y_{1i}^{1}(\zeta,\overline{\zeta
})-6\xi^{ij}(\tau)Y_{2ij}^{1}(\zeta,\overline{\zeta})+...\label{exp 3}\\
\psi_{2}^{0\,}  & \equiv\psi_{2(0,-3)}^{0\,}=\Upsilon+\psi_{2i}^{0}Y_{1i}%
^{0}+\psi_{2ij}^{0}Y_{2ij}^{0}+...\label{exp 4}\\
\psi_{1}^{0}  & \equiv\psi_{1(1,-3)}^{0}=\psi_{1i}^{0}Y_{1i}^{1}+\psi
_{1ij}^{0}Y_{2ij}^{1}+...\label{exp 5}\\
\Psi & \equiv\Psi_{(0,-3)}^{\,}=\Psi^{0}+\Psi^{i}Y_{1i}^{0}+\Psi^{ij}%
Y_{2ij}^{0}+...\label{exp 6}%
\end{align}
\begin{align}
\phi_{0}^{0}  & =\phi_{0i}^{0}Y_{1i}^{1}+\phi_{0ij}^{0}Y_{2ij}^{1}%
+...\label{harmonic decomposition}\\
\phi_{1}^{0}  & =Q+\phi_{1i}^{0}Y_{1i}^{0}+\phi_{1ij}^{0}Y_{2ij}^{0}+...\\
\phi_{2}^{0}  & =\phi_{2i}^{0}Y_{1i}^{-1}+\phi_{2ij}^{0}Y_{2ij}^{-1}+...
\end{align}

The basic idea is to try to give physical meaning or significance to the
harmonic coefficients. An important well known example of this is Bondi's
identification of the $l=(0,1)$ parts of the mass aspect with the total
energy-momentum four-vector of the interior sources, ($Mc,P^{i}$) which is
explicitly given by%

\begin{equation}
\Psi=-\frac{2\sqrt{2}G}{c^{2}}M-\frac{6G}{c^{3}}P^{i}Y_{1i}^{0}%
+....\label{M,P}%
\end{equation}

From the $l=0$ harmonic coefficient in Eq.(\ref{2*}) one obtains the Bondi
mass loss equation which allows us to identify the coefficient $\xi^{ij}(u)$
as proportional to the 2$^{nd}$ derivative of the mass quadrupole.

A few other identifications coming from Maxwell theory are: $Q\ $\ is the
Coulomb charge, while $\phi_{0i}^{0}\ $\ is proportional to the complex
electromagnetic dipole moment (electric $+$ $i$ magnetic), $\phi_{2i}^{0}$ is
proportional to the 2$^{nd}$ time derivative of $\phi_{0i}^{0}$ while
$\phi_{2ij}^{0}$ is proportional to the 3$^{rd}$ derivative of the quadrupole moment.

\textit{Very roughly speaking }we make the \textit{approximate} identification
of $\psi_{1i}^{0},$ (which is complex) with the complex gravitational dipole
moment, i.e., with the mass dipole plus `$i$' times the angular momentum.

\section{Identifications}

Our starting point for the physical identifications is Eq.(\ref{1})
\begin{equation}
(\psi_{1}^{0\,})^{\cdot}=-\text{$\eth $}\psi_{2}^{0}+2\sigma\text{$\eth $%
(}\overline{\sigma})^{\cdot}\label{1**}%
\end{equation}
and the spherical harmonic decomposition. We noted that the $l=1$ part of
$\psi_{2}^{0}$ was proportional to the Bondi three-momentum, $\vec{P}.$ The
linearized version of Eq.(\ref{1**}) leads to the fact that the $l=1$ part of
($\psi_{1i}^{0\,})^{\cdot}$ \ is also proportional to $\vec{P}.$ This in turn
suggests that $\psi_{1i}^{0\,}$ itself should be, at least in the linearized
version, proportional to the mass-dipole, $M\vec{X},$ where $\vec{X}$ should
be identified with the position vector of center of mass so that $M\vec
{X}^{\,\cdot}=\vec{P}.$

\textit{This is our fundamental observation. It will be analyzed and
generalized.}

We begin by postulating three different models or methods for the
identification. Though we have one method (the third) that we consider to be
fundamental and correct, nevertheless we felt that at least a few others
should be explored to see if they would be reasonable choices. \ Our criteria
for selecting a model are: (i) it should predict already known laws or
reasonable new laws, (ii) it should have a clear geometric foundation and have
a logical consistency and (iii) it should agree with special cases, mainly the
algebraically special metrics or analogies with flat-space Maxwell theory.

The first (naive) approach is to simply assume that the Bondi 3-momentum is
given by $\vec{P}=M\vec{X}^{\,\cdot}=M\overrightarrow{V}.^{\,}$The second
approach is based on the flat-space transformation (translation) properties of
dipoles, i.e., transforming to the center of mass or charge, while the third
approach generalizes this to asymptotically flat space-times. The first two do
not satisfy our three criteria while the third one does. We nevertheless felt
it was worthwhile to see how the increasing sophistication of the methods led
to improved results.

\textbf{Remark 1. }\textit{We have (for notational simplicity) totally abused
standard notation. We allow the indices (i,j,k,l...), which are Euclidean, to
be raised and lowered with impunity and allow repeated indices, e.g., }%
$v^{k}\xi^{ik},$ \textit{to indicate summation}$.$

\textbf{Remark 2. }\textit{Though u is the conventional Bondi time, it is more
appropriate to use }$w=\sqrt{2}u,$ \textit{it being the retarded time.
Derivatives with respect to u are denoted by dot, }($^{\cdot}$)\textit{, while
w derivatives are given by a prime, }($^{\prime}$)\textit{, i.e., }($^{\cdot}%
$)$=\sqrt{2}$($^{\prime}$).

\textbf{Claim} \textit{ A very important computational issue is to
find the inversion of Eq.(\ref{X*} ), i.e., to find
$\tau=T(u,\zeta,\overline{\zeta}).$ The reason for its importance is
that we must be able to explicitly eliminate the $\tau$ and replace
it by the Bondi $u$ (or $w$) in\ the parametric expressions for
$L(u,\zeta,\overline{\zeta}),$ i.e., in the expression for the
complex world-line $\xi^{a}(\tau)$. The approximate inversion
(linear) is given \cite{UCF}by}
\begin{equation}
\tau=T(u,\zeta,\overline{\zeta})=w+\frac{\sqrt{2}}{2}\xi^{i}(w)Y_{1i}%
^{0}(\zeta,\overline{\zeta})-\sqrt{2}\xi^{ij}(w)Y_{1ij}^{0}(\zeta
,\overline{\zeta})+....\label{inverse}%
\end{equation}

\subsection{\noindent The First Identification Method}

For our first (toy) model, we \textit{assume} that the Bondi momentum (at
least at linear order) has the standard kinematic form, namely
\begin{equation}
\vec{P}=M\,\vec{V},\label{MV}%
\end{equation}
and try to find the consequences. From our perspective this already is
unsatisfactory, since we would like this kinematic expression to follow
directly from the Einstein equations rather than to assume it. In fact, in the
second and third models this relationship is a derived result. Nevertheless
from a heuristic point of view we believe it worthwhile to show the line of
reasoning $-$ in this simpler case, - before is used later for the preferred
models. In addition, roughly speaking, from this first model we can see what
consequences can be anticipated.

Thus from the above argument, we take $\psi_{1i}^{0\,}$ to have the form%

\begin{equation}
\psi_{1i}^{0}=\alpha M\lambda_{i}\label{assumption1}%
\end{equation}
where $\alpha$ is a constant to be determined, $M$ is the Bondi mass and
$\lambda_{i}$ is a \textit{complex} three-vector in some unknown space
(hopefully) to be determined. The real part of $\lambda_{i}$ is tentatively
associated with the `center of mass position' and its derivative with the
`center of mass' velocity, i.e., $\lambda_{Ri}^{\prime}=v_{i}\Rightarrow
\overrightarrow{V}.$

Using the harmonic expansions, Eqs.(\ref{exp 4}), (\ref{exp 5}) and
Clebsch-Gordon products, [see appendix C], Eq.(\ref{1**}) becomes
\begin{equation}
\psi_{2k}^{0}=\frac{\sqrt{2}}{2c}\psi_{1k}^{0\prime}-i\frac{36\cdot64}%
{5c}\overline{\xi}^{mj\,\prime}\xi^{lm}\epsilon_{ljk}.\label{1***}%
\end{equation}
while the expansion of the mass aspect, Eq.(\ref{exp 6}), is
\begin{equation}
\Psi=\Upsilon(w)+\frac{16\cdot36\sqrt{2}}{5}\xi^{ij}\overline{\xi}%
^{ij\,\prime}+(\psi_{2k}^{0}+\frac{i(32)(36)}{5}\xi^{lj}\overline{\xi
}^{ij\,\prime}\epsilon_{ilk})Y_{1k}^{0}+...\label{exp 6*}%
\end{equation}
Using Eqs.(\ref{6*}), (\ref{1***}) and (\ref{assumption1}) we identify, from
Eq.(\ref{M,P}), the mass and three-momentum%

\begin{align*}
M  & =-\frac{c^{2}}{2\sqrt{2}G}\Upsilon-\frac{288}{5}\frac{c}{G}\xi
^{ij}\overline{\xi}^{ij\,\prime}\\
P^{k}  & =M\lambda_{k}^{\prime}+i\frac{192}{5G}c^{3}\overline{\xi}
^{jm\,\prime}\xi^{lm}\epsilon_{ljk}\\
\alpha & =-\frac{12G}{\sqrt{2}c^{2}}.
\end{align*}
Writing the variables as
\begin{align*}
\Upsilon & =\Upsilon_{R}+i\Upsilon\\
\lambda_{k}  & =\lambda_{Rk}+i\lambda_{Ik}\\
\xi^{lm}  & =\xi_{R}^{lm}+i\xi_{R}^{lm}\\
\xi^{lm\,\prime}  & \equiv v^{lm}=v_{R}^{lm}+iv_{R}^{lm}%
\end{align*}
the reality conditions, Eq.(\ref{3*}), becomes%

\begin{align}
M  & =-\frac{c^{2}}{2\sqrt{2}G}\Upsilon_{R}-\frac{36\cdot8}{5}\frac{c}{G}
(\xi_{R}^{ij}v_{R}^{ij\,}+\xi_{I}^{ij}v_{I}^{ij\,})\nonumber\\
M  & =M_{0}-\frac{36\cdot4}{5}\frac{c}{G}(\xi_{R}^{ij}\xi_{R}^{ij}+\xi
_{I}^{ij}\xi_{I}^{ij})^{\prime}\label{M}\\
\Upsilon_{I}  & =\frac{(24)^{2}\sqrt{2}}{5c}(\xi_{R}^{ij}v_{I}^{ij}-\xi
_{I}^{ij}v_{R}^{ij})\label{Y_I}\\
P^{k}  & =M\lambda_{Rk}^{\prime}-\frac{(24)^{2}c^{2}}{5G}(\xi_{I}^{il}%
v_{R}^{ij}-\xi_{R}^{il}v_{I}^{ij})\epsilon_{ljk}\label{P}\\
M\lambda_{Ik}^{\prime}  & =-\frac{(24)^{2}c^{2}}{5G}(\xi_{R}^{il}v_{R}%
^{ij}+\xi_{I}^{il}v_{I}^{ij})\epsilon_{ljk}\label{lambda_I}%
\end{align}

Finally, Eq.(\ref{2}),
\[
(\psi_{2}^{0\,})^{\cdot}=-\text{$\eth $}^{2}\text{(}\overline{\sigma})^{\cdot
}-\sigma(\overline{\sigma})^{\cdot\cdot},
\]
after using the Clebsch-Gordon expansion for $\sigma(\overline{\sigma}%
)^{\cdot\cdot},$ yields for the $l=0,1$ terms, the mass and momentum loss
expressions,
\begin{align}
M^{\prime}  & =-\frac{288c}{5G}(v_{R}^{ij}v_{R}^{ij}+v_{I}^{ij}v_{I}%
^{ij})\label{m'}\\
P^{k\prime}  & =F^{k}\equiv\frac{192c^{2}}{5G}(v_{I}^{lj}v_{R}^{ij}-v_{R}%
^{lj}v_{I}^{ij})\epsilon_{ilk}.\label{P^i'}%
\end{align}

There are now two identifications that can be made immediately from these
expressions. From the quadrupole energy loss relations\cite{LL}
\begin{equation}
M^{\,\prime}=-\frac{G}{5c^{7}}{\Large (}Q_{Mass}^{ij\,\prime\prime\prime
}Q_{Mass}^{ij\,\prime\prime\prime}+Q_{Spin}^{ij\,\prime\prime\prime}%
Q_{Spin}^{ij\,\prime\prime\prime}{\Large )}\label{M'*}%
\end{equation}
we can relate $\xi^{ij}$ to the mass and spin quadrupole moments by
\begin{equation}
\xi^{ij}=(\xi_{R}^{ij}+i\xi_{I}^{ij})=\frac{G}{12\sqrt{2}c^{4}}(Q_{Mass}%
^{ij\prime\prime}+iQ_{Spin}^{ij\prime\prime}).\label{lambda-Q}%
\end{equation}
This identification applies to all three models.

Furthermore, we can substitute the expression for $\overrightarrow{P},$ i.e.,
Eq.(\ref{P}), into the momentum loss equation, Eq.(\ref{P^i'}), obtaining
equations of motion for the `position' vector $\lambda_{R}^{i},$
\begin{equation}
M\lambda_{Rk}^{\prime\prime}-\frac{(24)^{2}c^{2}}{5G}(\xi_{I}^{ml}v_{R}%
^{mi}-\xi_{R}^{ml}v_{I}^{mi})^{\prime}\epsilon_{lik}=\frac{ 192c^{2}}%
{5G}(v_{I}^{lj}v_{R}^{ij}-v_{R}^{lj}v_{I}^{ij})\epsilon_{ilk}%
.\label{Eqs.Motion.1}%
\end{equation}

This can be interpreted as Newton's 2nd law, with the second term on the left
coming from gravitational radiation reaction and the right-side from the
`rocket' or recoil force due to momentum loss.

For the last identification in this model we consider Eq.(\ref{lambda_I}).
From the analogy with the rotating mass solution, (with and without charge
\cite{kerr-newman}), and results, which will be described later, we identify
\[
S_{k}=Mc\lambda_{Ik}
\]
as the spin angular momentum so that Eq.(\ref{lambda_I}) becomes the angular
momentum loss equation,%

\[
S_{k}^{\prime}=-\frac{(24)^{2}c^{3}}{5G}(\xi_{R}^{il}v_{R}^{ij}+\xi_{I}%
^{il}v_{I}^{ij})\epsilon_{ljk}.
\]

Summarizing the results for this physical model; we have \textit{assumed} that
the (complex) $l=1$ part of $\psi_{1k}^{0}$ was proportional to the mass
dipole moment and the spin angular momentum. The Bianchi identity Eq.(
\ref{1**}) allowed us to find a kinematic expression for the Bondi linear
momentum in terms of the mass dipole moment (or center of mass) while the
Bianchi identity, Eq.( \ref{2}), yielded the equations of motion. The $l=1$
reality condition became the angular momentum loss equation.

Though this model does lead to physical identifications that are quite
reasonable, it nevertheless contains several very unsatisfactory features.
First of all there is no hint of a geometric meaning for the `position' vector
$\lambda_{k}.$ What space is it in? It was pulled out of thin air, just so
that the momentum could be written as $\overrightarrow{P}=M \overrightarrow
{V}.$ The constant $\alpha$ was arbitrary until determined by the
\textit{\ kinematic meaning of \ }$\overrightarrow{P}.$ A second deficiency is
that the angular momentum is given only by the spin without any suggestion of
an orbital part. Lastly, this identification is at total odds with the
identifications obtained for the algebraically special metrics
\cite{RT,RTM,Talbot,UCF} or with the analogous flat space Maxwell
identification with the center of charge\cite{shearfreemax}.

We will see that the third model does correct these deficiencies.

\subsection{The Second Identification Method}

Our second and third identification models are based on complex curves, in the
second model these curves are associated with flat-space shear-free null
geodesic congruences while in the third model they are associated with the
asymptotically shear-free null geodesic congruences of the relevant
space-time. These curves, as we mentioned earlier, are geometrically well
defined as complex world-lines in the space of complex Poincare translations
at $\mathfrak{I}^{+}$. \ The ideas involved are not obvious and do lie outside
of the usual default understandings and thus it takes a bit of `getting used
to'. \ Basically, the idea is to generalize the standard action of moving the
origin of coordinates to a new point, (the center of mass), so that the mass
dipole vanishes for that origin. The generalization consists of making these
transformations (translations) complex and moving to the \textit{complex}
center of mass, so that not only does the mass dipole vanish there, but so
does the angular momentum. \ This type of procedure, which worked perfectly,
in linearized gravity and Maxwell theory\cite{gyro,Sh.Free.Max} must now be
applied to asymptotically flat space-times.

We begin with the observation that most attempts\cite{Szabados} at defining
the angular momentum have the imaginary part of the $l=1$\textit{\ } harmonic
of $\psi_{1}^{0}$ as part of the definition.

\textbf{Remark 3. }\textit{Most of the angular momentum definitions based on
symmetry arguments start with the l=1 part of }$\psi_{1}^{0\,}$ \textit{and
augment it with terms quadratic in the shear and its derivatives. Later in
this section we will return to the issue of these quadratic terms. See
Appendix 7.4.}

The central part of our argument is to use the $l=1$\textit{\ }part of the
null rotated Weyl component, $\psi_{1}^{0\ast\,},$ from Eq.(\ref{psi trans 1}%
), i.e.,
\begin{equation}
\psi_{1}^{0\ast\,}=\psi_{1}^{0\,}-3L\psi_{2}^{0\,}+3L^{2}\psi_{3}^{0\,}%
-L^{3}\psi_{4}^{0\,}\label{psi_1*}%
\end{equation}
as the transformed (translated) complex dipole moment and then set the $l=1$
\textit{\ }part of $\psi_{1}^{0\ast\,}$to zero to (partially) determine the
$L$ . (In flat space this type of operation \textit{really} moves the origin
of Bondi light-cones to a new complex origin in complex Minkowski space so
that the center of charge and the magnetic dipole both vanish.) In other
words, we will set $\psi_{1}^{0\ast\,}|_{l=1}=0$ so that the $l=1$ part of
$\psi_{1}^{0\,}$ is given by
\begin{equation}
\psi_{1}^{0\,\,}|_{l=1}=(3L\psi_{2}^{0\,}-3L^{2}\psi_{3}^{0\,}+L^{3}\psi
_{4}^{0\,})|_{l=1}\cong3(L\psi_{2}^{0\,})|_{l=1}...\label{psi_1i}%
\end{equation}
It is this relationship that had, in the first model, its counterpart in the
arbitrary assumption of Eq.(\ref{assumption1}). Now it has a justification and
a geometric meaning. Also it does not contain an arbitrary factor $\alpha.$
The quantity $\psi_{1}^{0\,\,}|_{l=1}$ is uniquely defined by the choice of
$L(u,\zeta,\overline{\zeta}).$

Our second and third models depend on two different, but related, choices of
$L(u,\zeta,\overline{\zeta}).$ For the third model, we choose the
$L(u,\zeta,\overline{\zeta})$ to determine an arbitrary \textit{asymptotically
shear-free null geodesic congruence}, via Eqs.(\ref{reg.sol}) and
(\ref{good cut}), while for the second model we choose the $L(u,\zeta
,\overline{\zeta})$ associated with \textit{flat-space shear-free null
geodesic congruences}, Eqs.(\ref{flat L}) and (\ref{flat X}). The
$L(u,\zeta,\overline{\zeta})$ in either case is determined by an arbitrary
complex world-line $\xi^{a}(\tau),$ which contains the information about the
center of mass world-line and the spin. Though the third model is more
consistent and has a more logical basis, for two reasons we first discuss
model two. It is easier to work with and the final results are very similar to
those of the third model.

Using the flat-space $L(u,\zeta,\overline{\zeta}),$ working with terms up to
2$^{nd}$ order, with frequent use of Clebsch-Gordon expansions, we find, from
Eq.(\ref{psi_1i}), that
\begin{equation}
\psi_{1i}^{0}=3\Upsilon[\xi^{i}(w)+i\frac{1}{2}\epsilon_{kji}v^{k}\xi
^{j}]+i\frac{3\sqrt{2}}{2}\epsilon_{lji}\psi_{2j}^{0}\xi^{l}-\frac{18}{5}
\psi_{2ij}^{0}\xi^{j}.\label{psi_1i .2}%
\end{equation}

There are two things to immediately notice: 1. since $\Upsilon$ is
proportional to the mass, the first term has exactly the same form as in the
first model with $\xi^{i}$ now replacing the $\lambda^{i}$ as the `complex
center of mass' and 2. the $\psi_{1i}^{0}$ is fully determined without the
constant $\alpha.$ In addition, new terms have appeared, some of which will
have immediate physical meaning. From the Bianchi Identity, Eq.(\ref{1}),
using Eq.(\ref{psi_1i .2}) and the reality conditions, we have for the $l=0$
component of $\psi_{2}^{0},$%

\begin{align*}
M  & =M_{0}-\frac{288c}{5G}(\xi_{R}^{ij}\xi_{R}^{ij\prime}+\xi_{I}^{ij}\xi
_{I}^{ij\prime})\\
M_{0}  & =-\frac{c^{2}}{2\sqrt{2}G}\Upsilon_{R}(w)\\
\Upsilon_{I}  & =\frac{24(24)\sqrt{2}}{5c}(\xi_{R}^{ij}\xi_{I}^{ij\prime}%
-\xi_{I}^{ij}\xi_{R}^{ij\prime})
\end{align*}
which are identical to those of the first model. However for the $l=1$
components new terms appear:

For the momentum $\overrightarrow{P}$ we have
\begin{align}
P^{k}  & =M_{0}v_{R}^{k}+\frac{M_{0}}{c}(v_{R}^{i}v_{I}^{j}-\xi_{I}^{i}%
v_{R}^{j\prime}-(\xi_{R}^{i}v_{I}^{j})^{\prime})\epsilon_{ijk}+R^{k}%
,\label{P**}\\
R^{k}  & =-\frac{3(24)c^{2}}{5G\sqrt{2}}(\xi_{R}^{i}\xi_{R}^{ik}+\xi_{I}%
^{i}\xi_{I}^{ik})^{\prime}-\frac{(24)^{2}c^{2}}{5G}(\xi_{I}^{il}\xi
_{R}^{ij\prime}-\xi_{R}^{il}\xi_{I}^{ij\prime})\epsilon_{ljk}.
\end{align}
The $l=1$ part of reality condition yields an expression that consists of two
types of terms, total time derivatives and the others. We define it as the
conservation of angular momentum equation:%

\begin{align}
J^{k\prime}  & =-\frac{(24)^{2}c^{3}}{5G}(\xi_{R}^{il}\xi_{R}^{ij\prime}%
+\xi_{I}^{il}\xi_{I}^{ij\prime})\epsilon_{ljk},\label{J'}\\
J^{k}  & =M_{0}c\xi_{I}^{k}+M_{0}(\xi_{R}^{i}v_{R}^{j}-\xi_{I}^{i}v_{I}%
^{j})\epsilon_{ijk}-\frac{3(24)c^{3}}{5\sqrt{2}G}(\xi_{I}^{i}\xi_{R}^{ik}%
-\xi_{R}^{i}\xi_{I}^{ik}).\label{J}%
\end{align}
The angular momentum flux is the same as in the first model. However the
angular momentum itself has much more physical content. The first term%

\begin{equation}
S^{k}=M_{0}c\xi_{I}^{k}\label{Spin}%
\end{equation}
is again identified as the intrinsic spin, while the second term is precisely
the orbital angular momentum, $\mathbf{r}\times\mathbf{p,}$ while the third
term is the contribution from spin-precession. The fourth term is new,
involving dipole and quadrupole coupling. This could be considered as a
prediction of the theory though it probably is untestable. With these
identifications, going back to $P^{k}$ in Eq.(\ref{P**})$,$ we observe that
the second term
\[
\frac{M}{c}\xi_{I}^{i\prime}v_{R}^{j}\epsilon_{ijk}=c^{-2}S^{i\prime}v_{R}%
^{j}\epsilon_{ijk}=c^{-2}(\mathbf{S}^{\prime}x\mathbf{v)}_{k}
\]
is the Mathisson-Papapetrou contribution to the linear momentum.

\textbf{Remark 4. }\textit{Many of the kinematic \ expressions for the
physical quantities have been given up to second order. When the dynamic
equations are used, it often turns out that these quantities are really of
higher order and should be neglected for second order considerations. We felt
that for purposes of understanding their kinematic role, their inclusion here
was important.}

The equations of motion for the position vector, $\xi_{R}^{k},$ are obtained
by substituting the $\overrightarrow{P}$ from Eq.(\ref{P**}) into the momentum
loss equation%

\[
P^{k\prime}=F^{k}\equiv\frac{192c^{2}}{5G} (v_{I}^{lj}v_{R}^{ij}-v_{R}%
^{lj}v_{I}^{ij})\epsilon_{ilk}
\]
yielding
\[
M_{0}v_{R}^{k\prime}-\frac{M_{0}}{c}(\xi_{R}^{i}v_{I}^{j\prime}+\xi_{I}%
^{i}v_{R}^{j\prime})^{\prime}\epsilon_{ijk}+R^{k\prime}=\frac{ 384c^{2}}%
{5G}v_{I}^{lj}v_{R}^{ij}\epsilon_{ilk}.
\]
This again is Newton's 2$^{nd}$ law but with spin-coupling forces, radiation
reaction and momentum recoil.

Though this second model is far superior to the first model it still has
certain deficiencies. It does not yield the correct dynamics for the
algebraically special metrics. In other words the dynamical equations would
differ from those of the algebraically special metrics. Also there is a lack
of logical consistency in this treatment. We have used the $L(u,\zeta
,\overline{\zeta})$ for the flat-space shear-free null geodesic congruence. We
should have used the $L(u,\zeta,\overline{\zeta})$ associated with the
\textit{asymptotically shear-free} congruence, i.e., the $L(u,\zeta
,\overline{\zeta})$ from Eq.( \ref{reg.sol}). We would then have the null
rotated Weyl component, $\psi_{1i}^{\ast0},$ based on an
\textit{\ asymptotically} \textit{shear-free} null congruence \textit{so that
the quadratic terms in the shear, that \textbf{would normally appear} in the
definition of angular momentum, would be now absent}, [See \textit{Remark} 3.]
and our procedure of setting $\psi_{1i}^{\ast0}=0$ would be consistent with
all definitions of center of mass and angular momentum at the complex `origin'.

\subsection{The Third Identification Method or Physical Model}

The third method is, as we mentioned earlier, basically the same as the second
method. So rather than repeat ourselves with the vacuum
`\textit{asymptotically} \textit{shear-free', }$L(u,\zeta,\overline{\zeta}),$
we will give the results for the asymptotically-flat Einstein-Maxwell
equations. Since the method and arguments are so similar to those of method
two and the detailed calculations are so long, we will simply summarize the results.

The physical identifications are substantially changed by the presence of the
Maxwell field. For the vacuum case our results apply to the general
asymptotically flat situation. For asymptotically flat GR, with the Maxwell
field, the situation is different. It turns out that the Weyl tensor condition
(\ref{psi_1i}) has a counterpart for the Maxwell tensor. In other words, the
$l=1$ part of a null rotated $\phi_{0}^{0}$, i.e., $\phi_{0i}^{0\ast}$ when
set to zero, determines a \textit{different complex world-line} $($the complex
center of charge) for which the electric and magnetic dipole moments vanish.
We consider only the \textit{special} case where the gravitational and
electromagnetic world-lines coincide. In some sense this coincidence of
world-lines implies that the source has a restricted structure and is
relatively `simple'. We take this as the \textit{meaning of a gravitational
elementary particle.}

We start with a very general cut function and its inverse that came from a
given Bondi shear, i.e., from Eqs.(\ref{good cut}) and (\ref{reg.sol}):
\begin{align}
u  & \equiv\frac{w}{\sqrt{2}}=X(\tau,\zeta,\overline{\zeta})=\frac{1}{\sqrt
{2}}\xi^{0}(\tau)-\frac{1}{2}\xi^{i}(\tau)Y_{1i}^{0}(\zeta,\overline{\zeta
})+\xi^{ij}(\tau)Y_{2ij}^{0}(\zeta,\overline{\zeta})+...,\label{X**}\\
\tau & =T(u,\zeta,\overline{\zeta})=w+\frac{\sqrt{2}}{2}\xi^{i}(w)Y_{1i}%
^{0}(\zeta,\overline{\zeta})-\sqrt{2}\xi^{ij}(w)Y_{1ij}^{0}(\zeta
,\overline{\zeta})+...\\
L  & =\xi^{i}(\tau)Y_{1i}^{1}(\zeta,\overline{\zeta})-6\xi^{ij}(\tau
)Y_{2ij}^{1}(\zeta,\overline{\zeta})\\
\xi^{a}  & =(\xi^{0},\text{ }\xi^{i}(\tau)=(\tau,\xi^{i}(\tau)=\xi_{R}%
^{a}(w)+i\xi_{I}^{a}(w)\qquad\\
\xi^{a\,\prime}  & \equiv v^{a}(w)=v_{R}^{a}(w)+iv_{I}^{a}(w)
\end{align}

After eliminating $\tau,$ Eq.(\ref{psi_1i}) yields the expression for
$\psi_{1i}^{0}:$%

\begin{align}
\psi_{1i}^{0}  & =3\Upsilon\lbrack\xi^{i}(w)+i\frac{1}{2}\epsilon_{kji}%
v^{k}\xi^{j}+N^{i}]+i\frac{3\sqrt{2}}{2}\epsilon_{lji}\psi_{2j}^{0}\xi
^{l}-\frac{18}{5}\psi_{2ij}^{0}\xi^{j}\label{dipole}\\
& -i\frac{6\cdot36\sqrt{2}}{5}\psi_{2kj}^{0}\xi^{kl\,}\epsilon_{lji}%
-\frac{6\cdot18}{5}\psi_{2j}^{0}\xi^{ij},\nonumber\\
N^{i}  & =\frac{6\sqrt{2}}{5}v^{k}\xi^{ki\,}-\frac{18\sqrt{2}}{5}v^{ki\,}%
\xi^{k}+i\frac{144}{5}\epsilon_{jmi}v^{kj\,}\xi^{mk\,}.
\end{align}

\qquad\textbf{Aside: }Later, for the vacuum case, we will associate the
imaginary part of Eq.(\ref{dipole}) as the total angular momentum.

\qquad Integrating the Maxwell equations leads to%

\begin{align}
\phi_{1i}^{0}  & =\sqrt{2}Q[v^{i}(w)+i\frac{1}{2}\epsilon_{ijl}v^{i\prime}%
\xi^{j}+N^{i\prime}]+i\sqrt{2}Q\epsilon_{kji}v^{j\prime}\xi^{k}-\frac{2}%
{5}(\phi_{0ki}^{0\prime}\xi^{k})^{\prime}-\frac{72}{5}Qv^{j}v^{ij\,}%
\label{max}\\
& +i\frac{24\sqrt{2}}{15}\epsilon_{lji}\phi_{0kj}^{0\prime\prime}\xi
^{lk\,}-i\frac{24\sqrt{2}}{5}\epsilon_{jli}(\phi_{0ml}^{0\prime}\xi
^{jm\,})^{\prime}\nonumber\\
\phi_{2i}^{0}  & =-2Q[v^{i\prime}+i\frac{1}{2}\epsilon_{ijl}(v^{i\prime}%
\xi^{j})^{\prime}+N^{i\prime\prime}]-i2Q\epsilon_{kji}(v^{j\prime}\xi
^{k})^{\prime}+\frac{2\sqrt{2}}{5}(\phi_{0ki}^{0\prime}\xi^{k})^{\prime\prime
}\nonumber\\
& +\frac{72\sqrt{2}}{5}Q(v^{j}v^{ij\,})^{\prime}-\frac{48}{15}i\epsilon
_{lji}(\phi_{0kj}^{0\prime\prime}\xi^{lk\,})^{\prime}+i\frac{48}{5}%
\epsilon_{jli}(\phi_{0ml}^{0\prime}\xi^{jm\,})^{\prime\prime}\nonumber
\end{align}

From the Bianchi Identity Eq.(\ref{Bondi1}) we obtain%

\begin{align}
P^{k}  & =Mv_{R}^{k}+\frac{M_{0}}{c}(v_{R}^{i}v_{I}^{j}-\xi_{I}^{i}%
v_{R}^{j\prime}-(\xi_{R}^{i}v_{I}^{j})^{\prime})\epsilon_{ijk}-\frac{2Q^{2}%
}{3c^{3}}v_{R}^{k{\Large \,}\prime}\label{P*}\\
& +\frac{2Q^{2}}{3c^{4}}[2\xi_{I}^{i}v_{R}^{j\,\prime}-\xi_{R}^{i}%
v_{I}^{j\,\prime}+v_{R}^{i}v_{I}^{j}]^{\prime}\epsilon_{ijk}+\Pi^{k}\\
\Pi^{k}  & =-\frac{M}{c}{\Large (}\frac{6\sqrt{2}}{5}[8(\xi_{R}^{ki}v_{R}%
^{i}-\xi_{I}^{ki}v_{I}^{i})+3(v_{R}^{ki}\xi_{R}^{i}-v_{I}^{ki}\xi_{I}%
^{i})]\,^{\prime}+\frac{144}{5} (v_{R}^{il\,\prime}\xi_{I}^{ij}+v_{I}%
^{il\,\prime}\xi_{R}^{ij})\epsilon_{ljk}{\Large )}\nonumber\\
& +\frac{Q^{2}}{3c^{4}}{\Large (}\frac{18(6)\sqrt{2}}{5}(v_{R}^{i\,\prime}%
\xi_{R}^{ik}+v_{I}^{i\,\prime}\xi_{I}^{ik})^{\,\prime}+\frac{96\sqrt{2}}%
{5}(v_{R}^{ki}v_{R}^{i}-v_{I}^{ki}v_{I}^{i})^{\,\prime}\nonumber\\
& -\frac{12\sqrt{2}}{5}(\xi_{R}^{ki}v_{R}^{i\,\prime}-\xi_{I}^{ki}%
v_{I}^{i\,\prime})^{\,\prime}+\frac{(36)\sqrt{2}}{5} (v_{R}^{ki\,\prime}%
\xi_{R}^{i}-v_{I}^{ki\,\prime}\xi_{I}^{i})^{\prime}+\frac{288}{5}%
(v_{R}^{il\,\prime}\xi_{I}^{ij}+v_{I}^{il\,\prime}\xi_{R}^{ij})^{\prime
}\epsilon_{ljk}{\Large )}\nonumber\\
& +\frac{Q}{3c^{4}}{\Large (}\frac{24}{5\sqrt{2}c}(D_{E}^{ij\,\prime\prime}%
\xi_{I}^{il}+D_{M}^{ij\,\prime\prime}\xi_{R}^{il})^{\prime\prime}%
\epsilon_{ljk}-\frac{1}{5c}(\xi_{R}^{i}D_{E}^{ik\,\prime\prime}-\xi_{I}%
^{i}D_{M}^{ik\,\prime\prime})^{\prime\prime}\nonumber\\
& -\frac{24}{15\sqrt{2}c}(D_{E}^{ij\,\prime\prime\prime}\xi_{I}^{il}%
+D_{M}^{ij\,\prime\,\prime\prime}\xi_{R}^{il})^{\prime}\epsilon_{ljk}-\frac
{1}{5c}(v_{R}^{i\,\prime}D_{E}^{ik\,\prime\prime}+v_{I}^{i\,\prime}%
D_{M}^{ik\,\prime\prime})+\frac{1}{5c} (v_{R}^{i}D_{E}^{ik\,\,\prime
\prime\prime}+v_{I}^{i}D_{M}^{ik\,\prime\prime\prime}){\Large )}\nonumber\\
& +\frac{4}{45c^{4}}(D_{M}^{ij\,\prime\prime}D_{E}^{il\,\prime\prime\prime
}-D_{E}^{ij\,\prime\prime}D_{M}^{il\,\prime\prime\prime})\epsilon_{ljk}%
-\frac{36\sqrt{2}\,c^{2}}{5G}(\xi_{R}^{i}\xi_{R}^{ik}+\xi_{I}^{i}\xi_{I}%
^{ik})^{\prime}\nonumber\\
& +\frac{2(24)^{2}c^{2}}{5G}(v_{R}^{ij}\xi_{I}^{il}+\xi_{R}^{ij}v_{I}%
^{il})\epsilon_{ljk}.\nonumber
\end{align}
$D_{E}^{lj\,}$ and $D_{M}^{ij\,}$ are respectively the electric quadrupole and
magnetic quadrupole moments found from the $l=2$ radiation term in the
solution of the Maxwell equations. All non-linear terms involving the
quadrupole terms are gathered into the $\Pi^{k}.$

The vanishing of the imaginary part of the reality condition yields the relations%

\begin{equation}
J^{k}{}^{\,\prime}=\frac{2Q^{2}}{3c^{3}}(v_{R}^{i\,\prime}v_{R}^{j}%
+v_{I}^{i\,\prime}v_{I}^{j})\epsilon_{ljk}+\frac{1}{90c^{5}}(D_{E}%
^{ij\,\prime\prime}D_{E}^{il\,\prime\prime\prime}+D_{M}^{ij\,\prime\prime
}D_{M}^{il\,\prime\prime\prime})\epsilon_{ljk}-\frac{(24)^{2}c^{3}}{5G}%
(\xi_{R}^{il}v_{R}^{ij}+\xi_{I}^{il}v_{I}^{ij})\epsilon_{ljk}\label{Jdot}%
\end{equation}
where $J^{k}$, identified (from the dynamics rather than through the
conventional symmetry argument) as the total angular momentum, is given by
\begin{align}
J^{k}  & \equiv Mc\xi_{I}^{k}+M(\xi_{R}^{i}v_{R}^{j}-\xi_{I}^{i}v_{I}%
^{j})\epsilon_{ijk}+\frac{2Q^{2}}{3c^{2}}v_{I}^{k}-\frac{2Q^{2}}{3c^{3}}%
(\xi_{R}^{i}v_{R}^{j\,\prime}+2\xi_{I}^{i}v_{I}^{j\,\prime})\epsilon
_{ijk}+K^{k}\label{J*}\\
K^{k}  & =-M{\Large (}\frac{6\sqrt{2}}{5}[8(\xi_{R}^{ki}v_{I}^{i}+\xi_{I}%
^{ki}v_{R}^{i})+3(v_{R}^{ki}\xi_{I}^{i}+v_{I}^{ki}\xi_{R}^{i})]-\frac{144}%
{5}(v_{R}^{il}\xi_{R}^{ij}-v_{I}^{il}\xi_{I}^{ij})\epsilon_{ljk}%
{\large )}\nonumber\\
& -\frac{Q^{2}}{3c^{3}}{\Large (}-\frac{18(6)\sqrt{2}}{5}(v_{R}^{i\,\prime}%
\xi_{I}^{ik}-v_{I}^{i\,\prime}\xi_{R}^{ik})+\frac{96\sqrt{2}}{5}(v_{R}%
^{ki}v_{I}^{i}+v_{I}^{ki}v_{R}^{i})\nonumber\\
& -\frac{12\sqrt{2}}{5}(\xi_{R}^{ki}v_{I}^{i\,\prime}+\xi_{I}^{ki}%
v_{R}^{i\,\prime})+\frac{(36)\sqrt{2}}{5}(v_{R}^{ki\,\prime}\xi_{I}^{i}%
+v_{I}^{ki\,\prime}\xi_{R}^{i})-\frac{288}{5}(v_{R}^{il\,\prime}\xi_{R}%
^{ij}-v_{I}^{il\,\prime}\xi_{I}^{ij})\epsilon_{ljk}{\Large )}\nonumber\\
& +\frac{Q}{15c^{4}}{\Large (}\xi_{R}^{i}D_{M}^{ik\,\prime\prime\prime}%
+\xi_{I}^{i}D_{E}^{ik\,\prime\prime\prime}+2v_{I}^{i}D_{E}^{ik\,\prime\prime
}+4\sqrt{2}(2\xi_{R}^{il}D_{E}^{ij\,\prime\prime\prime}-2\xi_{I}^{il}%
D_{M}^{ij\,\prime\prime\prime}\nonumber\\
& +3v_{R}^{il}D_{E}^{ij\,\prime\prime}-3v_{I}^{il}D_{M}^{ij\,\prime\prime
})\epsilon_{ljk}{\Large )}-\frac{36\sqrt{2}\,c^{3}}{5G}(\xi_{I}^{i}\xi
_{R}^{ik}-\xi_{R}^{i}\xi_{I}^{ik}).\nonumber
\end{align}
The justification for calling this the angular momentum is the same as in the
previous model, coming from the sum of terms interpretable as the spin,
orbital and precessional moments.

In the absence of a Maxwell field $J^{k}$ is simply proportional to the
imaginary part of $\psi_{1i}^{0},$%
\begin{equation}
J^{i}=-\frac{\sqrt{2}c^{3}}{12G}\psi_{1i}^{0}|_{I}.
\end{equation}

\textbf{Remark 5. }\textit{In the published literature\cite{dray} there are
ambiguities in the definition of angular momentum. For us, because of our
approximations, these ambiguities disappear, i.e., the ambiguous terms are
higher order. See Appendix 7.4.}

Our last results arise from are the mass/energy loss equation and the momentum
loss equation. The later being the dynamic equations for $\xi_{R}^{i}.$ By
substituting the $P^{k}$ of Eq.(\ref{P*}) into the Bianchi identity,
Eq.(\ref{Bondi3}) we find
\begin{align}
M^{\,\prime}  & =-\frac{G}{5c^{7}}{\Large (}Q_{Mass}^{ij\,\prime\prime\prime
}Q_{Mass}^{ij\,\prime\prime\prime}+Q_{Spin}^{ij\,\prime\prime\prime}%
Q_{Spin}^{ij\,\prime\prime\prime}{\Large )}-\frac{2Q^{2}}{3c^{5}}%
(v_{R}^{i\,\prime}v_{R}^{i\,\prime}+v_{I}^{i\,\prime}v_{I}^{i\,\prime
})\label{M'}\\
& -\frac{1}{180c^{7}}{\Large (}D_{E}^{ij\,\prime\prime\prime}D_{E}%
^{ij\,\prime\prime\prime}+D_{M}^{ij\,\prime\prime\prime}D_{M}^{ij\,\prime
\prime\prime}{\Large )}\nonumber\\
P^{k\,\prime}  & =F^{k}\equiv\frac{2G}{15c^{6}}{\Large (}Q_{Spin}%
^{lj\,\prime\prime\prime}Q_{Mass}^{ij\,\prime\prime\prime}-Q_{Mass}%
^{lj\,\prime\prime\prime}Q_{Spin}^{ij\,\prime\prime\prime}{\Large )}%
\epsilon_{ilk}-\frac{Q^{2}}{3c^{4}}(v_{I}^{l\,\prime}v_{R}^{i\,\prime}%
-v_{R}^{l\,\prime}v_{I}^{i\,\prime})\epsilon_{ilk}\label{P'}\\
& +\frac{Q}{15c^{5}}{\Large (}v_{R}^{j\,\prime}D_{E}^{jk\,\prime\prime\prime
}+v_{I}^{j\,\prime}D_{M}^{jk\,\prime\prime\prime}{\Large )}+\frac{1}{540c^{6}%
}{\Large (}D_{E}^{lj\,\prime\prime\prime}D_{M}^{ij\,\prime\prime\prime}%
-D_{M}^{lj\,\prime\prime\prime}D_{E}^{ij\,\prime\prime\prime}{\Large )}%
\epsilon_{ilk}\nonumber
\end{align}
where we have now replaced the $\xi^{ij}$ by the more physical variables, the
quadrupoles, ($Q_{Mass}^{ij\prime\prime},Q_{Spin}^{ij\prime\prime}$ ) via%

\begin{equation}
\xi^{ij}=(\xi_{R}^{ij}+i\xi_{I}^{ij})=\frac{G}{12\sqrt{2}c^{4}}(Q_{Mass}%
^{ij\prime\prime}+iQ_{Spin}^{ij\prime\prime}).\label{quadu}%
\end{equation}

Writing out the equations of motion in detail is long and not completely
enlightening, There are many non-linear terms whose meanings, other than they
are interpretable as gravitational radiation reaction, are not clear. Instead
we will write out a truncated version of Eq.(\ref{P'}), hiding these terms in
a single symbol, $R^{k},$ so that we have
\begin{equation}
Mv_{R}^{k\prime}+v_{R}^{k}M^{\prime}-\frac{2Q^{2}}{3c^{3}}v_{R}^{k{\Large \,}%
\prime\prime}+R^{k}=F^{k}.\label{P'*}%
\end{equation}
There are several things to notice here. First of all strictly speaking, we
should ignore the term $v_{R}^{k}M^{\prime}$ since it is third order because
$M^{\prime}$ already is second order from Eq.(\ref{M'}). We keep it,
understanding its suspect nature, because it is potentially so important. For
the moment ignore $R^{k}$and $F^{k}$ and consider only the second term in the
mass loss, i.e., the electric dipole energy loss, so that Eq.(\ref{P'*})
becomes
\begin{equation}
Mv_{R}^{k\prime}-v_{R}^{k}\frac{2Q^{2}}{3c^{5}}v_{R}^{i\,\prime}%
v_{R}^{i\,\prime}-\frac{2Q^{2}}{3c^{3}}v_{R}^{k{\Large \,}\prime\prime
}=0.\label{rad.react}%
\end{equation}
These\cite{Thirring,LL} are the classical equations of motion for a charged
particle that contain the well known electromagnetic radiation reaction force
and exhibit the unstable run-away behavior. (The cubic term, though not
sufficient large to stabilize the equation, has the correct sign for
stabilization.) This classical equation has been obtained \textit{without}
\textit{the usual} \textit{model building or mass renormalization. } Returning
to Eq.(\ref{P'*}), we see that in addition to the $R^{k}$and $F^{k},$ there
are the extra terms in $M^{\prime}$ coming from gravitational and
electromagnetic quadrupole radiation. It is very hard to see the consequences
these terms though one can see that they are on the side of stabilization. It
is then easy to conjecture that coupling electrodynamics with general
relativity stabilizes the equations of motion.

There are other comments concerning our results that should be made.

$\bullet$ The mass/energy loss equation contains both the gravitational
radiation quadrupole expression (this has been adjusted by definition) and the
classical electromagnetic dipole and quadrupole radiation expressions that
come straight from the construction. This allows us to identify
\begin{equation}
\mu^{k}=Q\xi_{I}^{k}\label{mm}%
\end{equation}
as the magnetic dipole moment, so that with the spin definition, $S^{k}%
=Mc\xi_{I}^{k},$ we obtain the Dirac value of the gyromagnetic ratio, i.e.,%

\begin{equation}
g=2.\label{g}%
\end{equation}

$\bullet$ The angular momentum expression, Eq.(\ref{J*}), contains the spin,
the orbital angular momentum and a precession contribution. There is now a
prediction that there is a charge-spin contribution to the total $J,$ i.e.,
$\frac{2Q^{2}}{3Mc^{3}}S^{k\prime},$ as well has higher order corrections.

$\bullet$ Probably the \textit{strongest argument} for the validity of this
approach to extracting physical information from the asymptotic field is the
observation that in the angular momentum loss equation, Eq.(\ref{Jdot}),
\textit{the dipole contribution} to the angular momentum flux coming just from
the electromagnetic field, i.e.,
\begin{equation}
\frac{2Q^{2}}{3c^{3}}(v_{R}^{i\,\prime}v_{R}^{j}+v_{I}^{i\,\prime}v_{I}%
^{j})\epsilon_{ljk}\label{di.con}%
\end{equation}
exactly coincides with the classical electrodynamic angular momentum flux
\cite{LL}. This result was obtained with no a priori expectations.

\section{BMS Invariance}

Our results concerning the identifications of physical quantities from the
asymptotic fields were all obtained in an arbitrary but specific Bondi
coordinate/tetrad system. The question is what are the relations between the
same quantities but calculated in a different Bondi coordinate/tetrad system?
In other words, we want to know the transformation properties of our physical
variables under the action of the BMS
group\cite{bondi,sachs,sachs2,penrose1,held-newman}. As we pointed out in Sec.
II, the BMS group is composed of two parts, the supertranslations and the
Lorentz transformations given respectively by
\begin{align}
\widehat{u}  & =u+\alpha(\zeta,\overline{\zeta})\label{supert*}\\
(\widehat{\zeta},\overline{\widehat{\zeta}})  & =(\zeta,\overline{\zeta
})\nonumber
\end{align}
with $\alpha(\zeta,\overline{\zeta})$ an arbitrary smooth function on the
sphere considered now to be small i.e., as a first order quantity and
\begin{align}
\widehat{u}  & =Ku\label{Lorentz*}\\
K  & =\frac{1+\zeta\overline{\zeta}}{(a\zeta+b)(\overline{a}\overline{\zeta
}+\overline{b})+(c\zeta+d)(\overline{c}\overline{\zeta}+\overline{d}%
)}\nonumber\\
\widehat{\zeta}  & =\frac{a\zeta+b}{c\zeta+d};\text{ \ }ad-bc=1.\nonumber
\end{align}
with $(a,b,c,d)$ the complex parameters of SL(2,C). The invariance under
supertranslations is actually subtle and requires a bit of thought while the
invariance under the Lorentz group, though straightforward, requires some more
technical background. Since all our calculations were done under the
assumption of second order perturbations off a Reissner-Nordstrom background,
we must keep the BMS transformations small, i.e., close to the identity.

\subsection{The Supertranslations}

We first treat supertranslation invariance.

Starting on $\mathfrak{I}^{+}$ with some arbitrary but given Bondi coordinates
$(u,\zeta,\overline{\zeta})$ and tetrad $(l,n,m,\overline{m})$ we saw that
there was a null rotation, Eq.(\ref{null rot}),
\begin{align}
l^{\ast}  & =l+L\overline{m}+\overline{L}m+L\overline{L}n\label{null rot*}\\
m^{\ast}  & =m+Ln\nonumber\\
n^{\ast}  & =n\nonumber
\end{align}
that was determined by the choice of the angle field $L(u,\zeta,\overline
{\zeta}).$ The $L(u,\zeta,\overline{\zeta})$ was then determined first by the
asymptotic shear-free condition, leaving the freedom in the $l=(0,1)$ harmonic
coefficients, and then by the requirement that the $l=1$ harmonic component of
the `rotated' Weyl component, $\psi_{1}^{0\ast\,},$ i.e., $\psi_{1}^{0\ast
\,}|_{l=1}=0,$ should vanish. The important point to note is that the Weyl
tensor component $\psi_{1}^{0\ast\,}$ is determined by the new tetrad,
$(l^{\ast},n^{\ast},m^{\ast},\overline{m}^{\ast}),$ It is a geometric
structure given independent of the choice of coordinates - depending on the
fixed known (*)-tetrad. If we have a second Bondi system with Bondi tetrad
$(\widehat{l},\widehat{n},\widehat{m},\overline{\widehat{m}}),$ it will have
been obtained by a different null rotation via some other angle field,
$\widehat{L},$ from the $(l,n,m,\overline{m})$ tetrad. This simply means that
there is a different null rotation, going now from the ( $\widehat{}$)-tetrad
to the previously determined (*)-tetrad. The $\psi_{1}^{0\ast\,}|_{l=1}$ thus
remains zero when this harmonic is extracted from the $\psi_{1}^{0\ast\,}$
$\ $when the `$u$' is held constant. There however is a serious issue that
must be raised. When we extracted the $l=1$ part\ of $\psi_{1}^{0\ast\,},$ it
was done at constant value of the Bondi `$u$'. But now, after the
supertranslation, the harmonic decomposition should be done at constant
`$\widehat{u}$'. However, it turns out, because of the first order BMS
supertranslation, $\widehat{u}=u+\alpha(\zeta,\overline{\zeta}),$ the
$\psi_{1}^{0\ast\,}|_{l=1}$ part of $\psi_{1}^{0\ast\,},$ now obtained by
holding `$\widehat{u}$' constant, remains zero up to second order. So up to
our accuracy, our results are supertranslation invariant. The more difficult
issue of how to deal with finite supertranslations will be discussed in the conclusion.

\subsection{Lorentz Transformations}

To deal with the Lorentz subgroup of the BMS group requires a review of the
representation theory.

The theory of the representations of the Lorentz group was beautifully
described by Gelfand, Graev and Vilenkin\cite{GGV} using homogeneous functions
of two complex variables as the representation space. We will summarize these
ideas using an equivalent method\cite{held-newman} namely by using
spin-weighted functions of the sphere as the representation spaces.
Representations are labeled by either of two numbers $(n_{1},n_{2})$ or
$(s,w),$ with $(n_{1},n_{2})=(w-s+1,w+s+1).$ The `$s$' is referred to as the
spin weight and `$w$' as the conformal weight. The representations are
referred to as $D_{(n_{1},n_{2})}.$ The special case of irreducible unitary
representations, which occur when $(n_{1},n_{2})$ are not integers, are not of
interest to us and will not be discussed. We will consider only the case when
$(n_{1},n_{2})$ are integers so that the $(s,w)$ then can take on integer or
half integer values. The representation space, for each $(s,w),$ are the
functions on the sphere, $\eta_{(s,w)}(\zeta,\overline{\zeta}),$ that can be
expanded in spin-weighted spherical harmonics,$_{\,\,s}Y_{lm}(\zeta
,\overline{\zeta}),$ so that
\begin{equation}
\eta_{(s,w)}(\zeta,\overline{\zeta})=\sum_{l=s}^{\infty}\eta_{(lm)\text{
\thinspace}\,\text{\thinspace}\,s}Y_{lm}(\zeta,\overline{\zeta}%
),\label{expansion}%
\end{equation}
and transform under the Lorentz group, Eq.(\ref{Lorentz*}), as%

\begin{equation}
\widehat{\eta}_{(s,w)}(\widehat{\zeta},\overline{\widehat{\zeta}%
})=e^{is\lambda}K^{w}\eta_{(s,w)}(\zeta,\overline{\zeta})\label{transforms}%
\end{equation}
with
\begin{align}
e^{i\lambda}  & =\frac{c\zeta+d}{\overline{c}\overline{\zeta}+\overline{d}%
},\label{lambda&K}\\
K  & =(1+\zeta\overline{\zeta})[(a\zeta+b)(\overline{a}\overline{\zeta
}+\overline{b})+(c\zeta+d)(\overline{c}\overline{\zeta}+\overline{d}%
)]^{-1},\nonumber
\end{align}
These representations, in general, are neither irreducible nor totally
reducible. For us the important point is that many of these representations do
possess an invariant finite-dimensional subspace which (often) corresponds to
the usual finite dimensional tensor representation space. Under the
transformation. (\ref{transforms}). the finite number of coefficients in these
subspaces transform among themselves. It is this fact which we heavily
utilize. \ More specifically we have two related situations: (1) when the
$(n_{1},n_{2})$ are both positive integers, (or $w$ $\eqslantgtr|s|$), there
will be a finite dimensional invariant subspace, $D_{(n_{1},n_{2})}^{+},$ and
(2) when the $(-n_{1},-n_{2})$ are both negative integers there will be an
\textit{infinite} dimensional invariant subspace, $D_{(-n_{1},-n_{2})}^{-}$.
One, however, can obtain a \textit{finite} dimensional representation for each
negative integer case by the following construction: One forms the factor
space, $D_{(-n_{1},-n_{2})}/D_{(-n_{1},-n_{2})}^{-}.$ This space is isomorphic
to one of the finite dimensional spaces associated with the positive integers.
The explicit form of the isomorphism, which is not needed here, is given in
Held \& Newman\cite{held-newman}.

Rather than give the full description of these invariant subspaces, which is
available elsewhere, we will confine ourselves to the few cases of relevance
to us.

I. For $s=0$ and $w=1,$ [$(n_{1},n_{2})=(2,2)$], the harmonics, $l=(0,1$) form
the invariant subspace. Applying this to the cut function, $X(\tau
,\zeta,\overline{\zeta}),$we obtain from%

\begin{align}
u  & =X_{(0,1)}=\frac{1}{\sqrt{2}}\xi^{0}(\tau)-\frac{1}{2}\xi^{i}(\tau
)Y_{1i}^{0}(\zeta,\overline{\zeta})+\xi^{ij}(\tau)Y_{2ij}^{0}(\zeta
,\overline{\zeta})+...\label{X_0,1}\\
\text{Invar. subspace}  & =\text{{}}\frac{1}{\sqrt{2}}\xi^{0}(\tau)-\frac
{1}{2}\xi^{i}(\tau)Y_{1i}^{0}(\zeta,\overline{\zeta})\label{I.S.}\\
\xi^{a}(\tau)  & =(\xi^{0}(\tau),\xi^{i}(\tau))=\text{Lorentz vector}%
\label{L.v.}%
\end{align}

This allows us to single out, in a Lorentz invariant manner, the four $l=(0,1
$) harmonic coefficients of the cut-function $X(\tau,\zeta,\overline{\zeta})$
as a complex position vector.

II. The mass aspect,
\begin{equation}
\Psi\equiv\Psi_{(0,-3)}^{\,}=\Psi^{0}+\Psi^{i}Y_{1i}^{0}+\Psi^{ij}Y_{2ij}%
^{0}+...\label{0,-3}%
\end{equation}
is an $s=0$ and $w=-3,$ [$(n_{1},n_{2})=(-2,-2)$] quantity. The factor space
is isomorphic to the finite dimensional positive integer space, [$(n_{1}%
,n_{2})=(2,2)$] and hence the harmonics coefficients of $l=(0,1$) lie in the
invariant subspace. \ From the isomorphism, (which does change the numerical
coefficients) we can construct functions of the form Eq.(\ref{I.S.} ), which,
in turn, lead to
\[
P^{a}=(Mc,P^{i})=\text{Lorentz vector}
\]

This gives the justification for calling the $l=(0,1$) harmonics of the mass
aspect, a Lorentzian four-vector, $P^{a}$.

III. The Weyl tensor component, $\psi_{1}^{0},$ has $s=1$ and $w=-3,$
$[(n_{1},n_{2})=(-3,-1)].$ The associated finite dimensional factor space is
isomorphic to the finite part of the $s=-1,w=1,[(n_{1},n_{2})=(3,1)]$
representation. We have that
\begin{equation}
\psi_{1}^{0}\equiv\psi_{1(1,-3)}^{0}=\psi_{1i}^{0}Y_{1i}^{1}+\psi_{1ij}%
^{0}Y_{2ij}^{1}+...\label{1,-1}%
\end{equation}
leads to the invariant subspace%

\[
\text{Invariant subspace}=\text{{}}\psi_{1i}^{0}Y_{1i}^{-1}.
\]
The question of what finite tensor transformation does this correspond to is
slightly more complicated than that of the previous examples of Lorentzian
vectors. In fact it corresponds to the Lorentz transformations applied to
(complex) self-dual antisymmetric two-index tensors. As an example from
Maxwell theory, from a given \textbf{E} and \textbf{B, } the Maxwell tensor,
$F^{ab},$ and then its self-dual version
\[
W^{ab+}=F^{ab}+iF^{*ab}
\]
can be constructed. A Lorentz transformation applied to $W^{ab+}$ is
equivalent\cite{LL}, (see appendix B) to the same transformation applied to
\begin{equation}
\psi_{1i}^{0}=(\mathbf{E}+i\mathbf{B)}_{i}.\label{E+iB}%
\end{equation}

These observations allow us to assign invariant physical meaning to our
identifications of the position vector, $\xi^{a},$ the Bondi momentum, $P^{a}$
and the angular momentum $J^{i}.$

\section{Discussion}

Starting from a very unorthodox point of view, we have tried to describe in
the context of GR, (either in the vacuum case or for the Einstein-Maxwell
equations), the equations of motion of an \textit{isolated} charged, massive
body (\textit{our gravitational elementary particle}) that possesses both an
intrinsic spin and quadrupole moments and can radiate both gravitational and
electromagnetic radiation. The point of view arises from considering a
generalization of the algebraically special metrics where one can identify
physical quantities and determine their evolution from the asymptotic field equations.

For the algebraically special metrics, the shear-free null geodesic congruence
that are associated with these metrics, \textit{automatically} assigned
kinematic variables (e.g., a position vector and an intrinsic spin) to the
Bondi energy-momentum four-vector. The Bondi evolution equations then became
the equations of motion. Our generalization consisted in observing that the
existence of a \textit{shear-free null geodesic congruence} could be
generalized to the existence of an \textit{asymptotically shear-free null}
\textit{geodesic congruence}. Applying the same physical identifications as
with the shear-free congruences leads in exactly the same manner to the more
general equations of motion.

To test out if this result was accidental and, perhaps, other physical
assignments would lead to similar equations of motion, we tried two alternate
strategies. They both were found to be lacking. They were not as natural or as
physically meaningful as the third identification method.

Since we are far from any of the standard or default approaches to the
description of motion in GR it would be appropriate to summarize our results.

Looking at the asymptotic Einstein-Maxwell equations, the asymptotic
shear-free conditions lead, in general, to two different complex world-lines
in the space of complex Poincare translations acting on $\mathfrak{I}^{+},$
one from the Weyl tensor, the other from the Maxwell tensor. We have
considered only the special case where the two world-lines coincide - defining
this case as ``elementary particles in $GR".$

Some results among others are:

$\bullet$ The mass has a kinematic correction term dependent on the variable
quadrupole moment. This could perhaps be considered as a prediction.

$\bullet$The Bondi linear three-momentum is expressed in kinematic variables,
e.g., $Mv^{k}$, $\frac{2Q^{2}}{3c^{3}}v_{R}^{k\,\prime}$ and the
Mathisson-Papapetrou spin coupling, among others.

$\bullet$ The imaginary part of the complex position vector is identified with
the specific intrinsic spin angular momentum. From the solutions to Maxwell's
equations the magnetic moment is seen to be the charge, $q$, times the
imaginary part of the position vector. This agrees with the algebraically
special charged spinning metric and leads to the Dirac value of the
gyromagnetic ratio, i.e., $g=2.$ Though earlier we have defined a
`gravitational elementary particle' from this result, it should be noted that
in the elementary particle community\cite{Tony} it has been speculated that
all charged elementary particles with spin have this property. One then has
the question, what if any, is the relationship between the two types of particle?

$\bullet$ One of the strongest arguments for our interpretations comes from
the $l=1$ part of the reality conditions, which is interpreted as the dynamics
(the conservation law) for the total angular momentum. There is a total time
derivative term of a quantity, $J^{k},$ that we define as the total angular
momentum. It contains the spin, the orbital angular momentum and a precession
term. The angular momentum flux contains three terms which come, respectively,
from the gravitational quadrupole radiation, the electromagnetic quadrupole
radiation and a term arising from the electromagnetic (electric and magnetic)
dipole radiation. This latter term is identical to that calculated purely from
electromagnetic theory\cite{LL}.

$\bullet$ From the Bondi mass loss equation, we can identify from the flux
terms, the gravitational quadrupole but also see that our identification of
the electromagnetic dipole moments agrees with the predicted dipole energy loss.

$\bullet$ From the Bondi momentum loss we obtain the equations of motion. In a
sense we \textquotedblleft derive\textquotedblright\ Newton's 2$^{nd}$ law ,
$F=Mv,$ where the force is a combination of electromagnetic radiation
reaction\cite{Thirring,LL}, gravitational radiation reaction and a
\textquotedblleft rocket\textquotedblright\ recoil force from the
electromagnetic and gravitational momentum loss.

$\bullet$ Finally, we showed that each of the quantities that were identified
as physical variables transformed appropriately under the Lorentz group, i.e.,
as Lorentzian tensorial objects.

$\bullet$ Our results follow from the existence of a well-defined geometric
structure, namely the UCF, a unique one-complex parameter family of slices of
null infinity. This suggests that the higher order coefficients in the
harmonic expansion of the UCF, e.g., $\xi^{ij},$ $\xi^{ijk},...,$ \ should be
identified with time-derivatives of the higher multipole moments.

There are other unfamiliar terms that could be thought of as predictions of
this theoretical construct. How to possibly measure them is not at all clear.

$\bullet$ \ One interesting physical prediction concerns the contribution that
the charge makes to the total angular momentum. Looking at the equation
defining $J^{k},$ Eq.(\ref{J*}), we see the (linear) contribution from
\[
\frac{2Q^{2}}{3c^{2}}v_{I}^{k}=\frac{2Q}{3c^{2}}D_{M}^{k\prime},
\]
which is a coupling between the charge and the changing magnetic dipole moment.

The final item to be discussed concerns the invariance under the
supertranslation subgroup of the BMS group.

An obvious question concerning the material described here concerns the issue
of the extraction of the $l=(0,1)$ harmonics from different Weyl tensor
components or from the universal cut function, $X$. We have consistently
performed the extraction of the harmonic components on the cuts, $u=const.,$
or on neighboring cuts, $\widehat{u}=const.+\Delta,$ that are close to the
first set of cuts. To 2$^{nd}$ order the results are unchanged. If we do go to
arbitrary cuts, there is no reason for the extraction to lead to the same
results. Our results thus appear to depend on the choice of cuts. In fact
there is a canonical choice of cuts, i.e., a special one-parameter family of
cuts, labeled by `$s$', on which the extraction should always be performed and
for which there is no ambiguity. In the text, the $u=const.$ cuts were
sufficiently close to the canonical choice, so that, to 2$^{nd}$ order, they
were the same. In fact, in principle, we should have been doing all our
calculations on the $s=const.$ cuts. It was, however, easier doing it with
$u=const.$

The question then is what are these canonical cuts? Returning to the complex
universal cut-function
\[
u=X(\tau,\zeta,\overline{\zeta})
\]
we saw earlier that the $\tau$ had to be chosen so that the $u$ had real
values. If we write%

\begin{equation}
\tau=s+i\Lambda(s,\zeta,\overline{\zeta})\label{s+ilambda}%
\end{equation}
then one can show \{see appendix A\} that $\Lambda(s,\zeta,\overline{\zeta})$
can be chosen so that
\begin{equation}
u=X(s+i\Lambda(s,\zeta,\overline{\zeta}),\zeta,\overline{\zeta})=\widehat
{X}(s,\zeta,\overline{\zeta})\label{real slices}%
\end{equation}
is a real function of the real variable `$s$'$.$ This is the construction of
our canonical slicing.

\section{Acknowledgments}

G.S.O. acknowledges the financial support from CONACYT and Sistema Nacional de
Investigadores (SNI-M\'{e}xico). C.K. thanks CONICET and SECYTUNC for support.

\section{Appendix}

\subsection{The Canonical Slicing}

To construct the canonical slicing we begin with the complex UCF
\begin{align}
u  & =X(\tau,\zeta,\overline{\zeta})\\
& =\frac{1}{\sqrt{2}}\xi^{0}(\tau)-\frac{1}{2}\xi^{i}(\tau)Y_{1i}^{0}%
(\zeta,\overline{\zeta})+\xi^{ij}(\tau)Y_{2ij}^{0}(\zeta,\overline{\zeta
})+...\nonumber
\end{align}
and write
\begin{equation}
\tau=s+i\lambda\label{s+ilambea2}%
\end{equation}
with $s$ and $\lambda$ real. The cut function can then be rewritten
\begin{align}
u  & =X(\tau,\zeta,\overline{\zeta})=X(s+i\lambda,\zeta,\overline{\zeta})\\
& =\chi_{R}(s,\lambda,\zeta,\overline{\zeta})+i\chi_{I}(s,\lambda
,\zeta,\overline{\zeta}),\nonumber
\end{align}
with real $\chi_{R}(s,\lambda,\zeta,\overline{\zeta})\ $and $\chi
_{I}(s,\lambda,\zeta,\overline{\zeta}).$ The $\chi_{R}(s,\lambda
,\zeta,\overline{\zeta})\ $and $\chi_{I}(s,\lambda,\zeta,\overline{\zeta})$
are easily calculated from $X(\tau,\zeta,\overline{\zeta})$ by
\begin{align}
\chi_{R}(s,\lambda,\zeta,\overline{\zeta})  & =\frac{1}{2}\{X(s+i\lambda
,\zeta,\overline{\zeta})+\overline{X(s+i\lambda,\zeta,\overline{\zeta})}\}\\
\chi_{I}(s,\lambda,\zeta,\overline{\zeta})  & =\frac{1}{2}\{X(s+i\lambda
,\zeta,\overline{\zeta})-\overline{X(s+i\lambda,\zeta,\overline{\zeta}%
)}\}.\nonumber
\end{align}
By setting
\begin{equation}
\chi_{I}(s,\lambda,\zeta,\overline{\zeta})=0
\end{equation}
and solving for
\begin{equation}
\lambda=\Lambda(s,\zeta,\overline{\zeta})
\end{equation}
we obtain the real slicing,
\begin{equation}
u=\chi_{R}(s,\Lambda{\small (}s,\zeta,\overline{\zeta}{\small )}%
,\zeta,\overline{\zeta}).
\end{equation}

\textbf{Remark 6. \ }\textit{We remark without proof\cite{shearfreemax}, that
using the gauge freedom described early for the choice of the parameter }%
$\tau$, \textit{we can normalize the real velocity vector, }$v_{R}^{a}%
(s)=\xi_{R}^{a\prime}(s=\tau),$ \textit{to one, i.e., }$\eta_{ab}v_{R}%
^{a}(s)v_{R}^{b}(s)=1.$

\subsection{\noindent\ Lorentzian Tensors}

In Sec. IV we pointed out that certain spin and conformal weighted function on
the sphere carried finite dimensional representations of the Lorentz group,
i.e., they carried information about Lorentzian tensor objects. As examples of
this we will work out two specific cases.

Starting with the Lorentz transformation
\begin{align}
\widehat{\zeta}  & =\frac{a\zeta+b}{c\zeta+d},\qquad ad-bc=1\\
e^{i\lambda}  & =\frac{c\zeta+d}{\overline{c}\overline{\zeta}+\overline{d}}\\
K  & =\frac{(1+\zeta\overline{\zeta})}{[(a\zeta+b)(\overline{a}\overline
{\zeta}+\overline{b})+(c\zeta+d)(\overline{c}\overline{\zeta}+\overline{d}%
)]}\\
\widehat{\eta}_{(s,w)}(\widehat{\zeta},\overline{\widehat{\zeta}})  &
=e^{is\lambda}K^{w}\eta_{(s,w)}(\zeta,\overline{\zeta})
\end{align}
we choose the special transformation
\begin{align}
\widehat{\zeta}  & =a^{2}\zeta\label{special}\\
e^{i\lambda}  & =\frac{d}{\overline{d}}=\frac{\overline{a}}{a}\nonumber\\
K  & =\frac{a\overline{a}(1+\zeta\overline{\zeta})}{[(a\overline{a})^{2}%
\zeta\overline{\zeta}+1]}\nonumber
\end{align}

\textbf{Example 1 \ \ }$s=0$ and $w=1$

Applying this special transformation to the invariant subspace of an $s=0$ and
$w=1$ quantity, e.g., to $u=X(\tau,\zeta,\overline{\zeta}),$ we have%

\begin{align}
\widehat{\eta}_{(0,1)}(\widehat{\zeta},\overline{\widehat{\zeta}})  &
=K\eta_{(0,1)}(\zeta,\overline{\zeta})\label{a}\\
\widehat{\xi}^{a}l_{a}(\widehat{\zeta},\overline{\widehat{\zeta}})  &
=K\xi^{a}l_{a}(\zeta,\overline{\zeta})\label{b}\\
l_{a}(\zeta,\overline{\zeta})  & =\frac{\sqrt{2}}{2}(1,\frac{\zeta
+\overline{\zeta}}{1+\zeta\overline{\zeta}},-i\frac{\zeta-\overline{\zeta}%
}{1+\zeta\overline{\zeta}},\frac{-1+\zeta\overline{\zeta}}{1+\zeta
\overline{\zeta}}).\label{c}%
\end{align}

After using Eq.(\ref{special}) in Eq.(\ref{b}) and comparing the coefficients
of $(1$,$\zeta,\overline{\zeta},\zeta\overline{\zeta}),$ we find that%

\begin{align}
\widehat{\xi}^{0}  & =\frac{1}{2}(a\overline{a}+a^{-1}\overline{a}^{-1}%
)\xi^{0}+\frac{1}{2}(a^{-1}\overline{a}^{-1}-a\overline{a})\xi^{3}%
\label{Lorentz0}\\
\widehat{\xi}^{3}  & =(\frac{1}{2}a\overline{a}+\frac{1}{2}a^{-1}\overline
{a}^{-1})\xi^{3}+(\frac{1}{2}a^{-1}\overline{a}^{-1}-\frac{1}{2}a\overline{a}
)\xi^{0}\label{Lorentz1}\\
\widehat{\xi}^{1}-i\widehat{\xi}^{2}  & =\frac{\overline{a}}{a}(\xi^{1}%
-i\xi^{2})\label{rot1}\\
\widehat{\xi}^{1}+i\widehat{\xi}^{2}  & =\frac{a}{\overline{a}}(\xi^{1}%
+i\xi^{2}).\label{rot2}%
\end{align}

Since $\frac{a}{\overline{a}},$can be written as $e^{i\varphi},$we have a
spatial rotation in the $(1,2)$ plane. \ Then by identifying
\begin{equation}
(1-\frac{v^{2}}{c^{2}})^{-\frac{1}{2}}=\frac{1}{2}(a\overline{a}%
+a^{-1}\overline{a}^{-1})\label{gamma}%
\end{equation}
we have the Lorentz transformation
\begin{align}
\widehat{\xi}^{0}  & =\frac{\xi^{0}}{(1-\frac{v^{2}}{c^{2}})^{\frac{1}{2}}%
}+\frac{\frac{v}{c}\xi^{3}}{(1-\frac{v^{2}}{c^{2}})^{\frac{1}{2}}},\\
\widehat{\xi}^{3}  & =\frac{\xi^{3}}{(1-\frac{v^{2}}{c^{2}})^{\frac{1}{2}}%
}+\frac{\frac{v}{c}\xi^{0}}{(1-\frac{v^{2}}{c^{2}})^{\frac{1}{2}}}.
\end{align}
We see that the special fractional linear transformation $\widehat{\zeta}
=a^{2}\zeta$ corresponds to the standard Lorentz transformation with a spatial rotation.

\textbf{Example 2 \ \ }$s=-1$ and $w=1$ coming from the $s=1$ and $w=-3$ isomorphism.

Applying $\widehat{\zeta}=a^{2}\zeta$ to the $s=-1,w=1,$ case, e.g., to the
invariant factor space of $\psi_{1}^{0}\equiv\psi_{1(1,-3)}^{0}=\psi_{1i}%
^{0}Y_{1i}^{1}+\psi_{1ij}^{0}Y_{2ij}^{1}+...$

\noindent we have
\begin{align}
\widehat{\eta}_{(-1,1)}  & =\widehat{\psi}_{1i}^{0}\overline{m}_{i}%
(\widehat{\zeta},\overline{\widehat{\zeta}})=e^{-i\lambda}K\eta_{(-1,1)}%
=e^{-i\lambda}K\psi_{1i}^{0}(\zeta,\overline{\zeta})\\
\overline{m}_{a}(\zeta,\overline{\zeta})  & =\frac{\sqrt{2}}{2P}(0,1-\zeta
^{2},-i(1+\zeta^{2}),\text{ }2\zeta).\nonumber
\end{align}
Comparing the coefficients of $(1,\zeta,\zeta^{2}),$ we find that
\begin{align}
\widehat{\psi}_{1,1}^{0}  & =\frac{1}{2}(a^{2}+a^{-2})\psi_{1,1}^{0}-i\frac{
1}{2}(a^{2}-a^{-2})\psi_{1,2}^{0}\label{LT3}\\
& =\frac{\psi_{1,1}^{0}}{(1-\frac{v^{2}}{c^{2}})^{\frac{1}{2}}}-i\frac
{\frac{v}{c}\psi_{1,2}^{0}}{(1-\frac{v^{2}}{c^{2}})^{\frac{1}{2}}},\\
\widehat{\psi}_{1,2}^{0}  & =\frac{1}{2}[a^{-2}+a^{2}]\psi_{1,2}^{0}+\frac
{1}{2}i[a^{2}-a^{-2}]\psi_{1,1}^{0}\nonumber\\
& =\frac{\psi_{1,2}^{0}}{(1-\frac{v^{2}}{c^{2}})^{\frac{1}{2}}}+i\frac
{\frac{v}{c}\psi_{1,1}^{0}}{(1-\frac{v^{2}}{c^{2}})^{\frac{1}{2}}},\\
\widehat{\psi}_{1,3}^{0}  & =\psi_{1,3}^{0}.\nonumber
\end{align}

If we had identified $\psi_{1i}^{0}$ with a Maxwell field via
\begin{equation}
\psi_{1i}^{0}=(\psi_{1,1}^{0},\psi_{1,2}^{0},\psi_{1,3}^{0})=(\mathbf{E}
+i\mathbf{B)}_{i},
\end{equation}
then Eq.(\ref{LT3}) would be equivalent to a Lorentz transformation of the
Maxwell tensor $F^{ab}.$ The six real components of $\psi_{1i}^{0}$ thus
corresponds to a skew-symmetric Lorentzian tensor\cite{LL}.

\subsection{ Products of Spin-s Harmonics}

For completeness we give several of the relevant Clebsch-Gordon products that
were used. We have left out terms with $l$-values greater than two.%

\begin{align}
Y_{2kl}^{2}Y_{2ij}^{-2}  & =\frac{\delta_{ik}\delta_{jl}}{5}+i\frac{\sqrt{2}%
}{5}\delta_{jl}\epsilon_{ike}Y_{1e}^{0}\text{ }-\frac{1}{7}\delta_{lj}%
Y_{2ik}^{0},\\
Y_{2kl}^{0}Y_{2ij}^{0}  & =\frac{24}{5}\delta_{ik}\delta_{jl}+\frac{24}%
{7}\delta_{lj}Y_{2ik}^{0},\nonumber\\
Y_{2kl}^{-1}Y_{2ij}^{0}  & =-\frac{i12\sqrt{2}}{5}\delta_{ik}\epsilon
_{ljf}Y_{1f}^{-1}+\frac{12}{7}\delta_{li}Y_{2kj}^{-1},\nonumber\\
Y_{2kl}^{1}Y_{2ij}^{0}  & =\frac{i12\sqrt{2}}{5}\delta_{ik}\epsilon
_{ljf}Y_{1f}^{1}+\frac{12}{7}\delta_{li}Y_{2kj}^{1},\nonumber\\
Y_{1i}^{1}Y_{1j}^{-1}  & =\frac{1}{3}\delta_{ij}-\frac{i\sqrt{2}}{4}%
\epsilon_{ijk}Y_{1k}^{0}-\frac{1}{12}Y_{2ij}^{0},\nonumber\\
Y_{2ij}^{2}Y_{1k}^{-1}  & =\frac{3}{5}\delta_{jk}Y_{1i}^{1}-\frac{i\sqrt{2}}{
6}\epsilon_{ikl}Y_{2jl}^{1}\text{ ,}\qquad\bullet\text{ } Y_{2ij}^{2}%
Y_{1k}^{0}=\frac{i2\sqrt{2}}{3}\epsilon_{ikl}Y_{2jl}^{2},\nonumber\\
Y_{2ij}^{2}Y_{2ml}^{0}  & =-\frac{24}{7}\delta_{lj}Y_{2im}^{2},\qquad
\bullet\text{ }Y_{2lm}^{2}Y_{2ij}^{-1}=\frac{2\sqrt{2}}{5}i\delta_{im}%
\epsilon_{ljf}Y_{1f}^{1}+\frac{6}{7}\delta_{li}Y_{2mj}^{1},\nonumber\\
\text{ }Y_{2ij}^{-1}Y_{1k}^{1}  & =\frac{3}{5}Y_{1i}^{0}\delta_{jk}%
+\frac{i\sqrt{2}}{6}\epsilon_{jkl}Y_{2il}^{0}\text{ ,}\qquad\text{ }%
\bullet\text{ }Y_{2ij}^{1}Y_{1k}^{-1}=\frac{3}{5}Y_{1i}^{0}\delta_{jk}%
-\frac{i\sqrt{2}}{6}\epsilon_{jkl}Y_{2il}^{0},\nonumber\\
\text{ }Y_{1k}^{1}Y_{1i}^{0}  & =\frac{i}{\sqrt{2}}\epsilon_{kil}Y_{1l}%
^{1}+\frac{1}{2}Y_{2ki}^{1}\text{ ,}\qquad\text{ }\bullet\text{ }Y_{1k}%
^{-1}Y_{1i}^{0}=-\frac{i}{\sqrt{2}}\epsilon_{kil}Y_{1l}^{-1}+\frac{1}%
{2}Y_{2ki}^{-1},\nonumber\\
\text{ }Y_{1k}^{1}Y_{2ij}^{0}  & =-\frac{6}{5}Y_{1j}^{1}\delta_{ik}+i\sqrt{2}
\epsilon_{kil}Y_{2jl}^{1}\text{ ,}\qquad\text{ }\bullet\text{ }Y_{1k}%
^{-1}Y_{2ij}^{0}=-\frac{6}{5}Y_{1j}^{-1}\delta_{ik}-i\sqrt{2}\epsilon
_{kil}Y_{2jl}^{-1},\nonumber\\
\text{ }Y_{2ml}^{1}Y_{1i}^{0}  & =\frac{6}{5}Y_{1l}^{1}\delta_{im}-\frac{i}%
{3}\sqrt{2}\epsilon_{imf}Y_{2lf}^{1}\text{ ,}\qquad\text{ }\bullet\text{ }
Y_{2kl}^{-1}Y_{1i}^{0}=\frac{6}{5}Y_{1l}^{-1}\delta_{ik}+\frac{i}{3}\sqrt{2}
\epsilon_{ikf}Y_{2lf}^{-1},\nonumber\\
\text{ }Y_{1i}^{0}Y_{1j}^{0}  & =\frac{2}{3}\delta_{ij}+\frac{1}{3}Y_{2ij}%
^{0}\text{ ,}\qquad\text{ }\bullet\text{ }Y_{1i}^{0}Y_{2jk}^{0}=\frac{12}%
{5}\delta_{ij}Y_{1k}^{0},\nonumber
\end{align}

\subsection{ Angular Momentum Ambiguities}

As we mentioned earlier in the text there have been ambiguities, described in
the literature\cite{dray}, in the definition of the asymptotic angular
momentum, $J^{\ast k}$. In our notation, (omitting the Maxwell field), the
ambiguities are in the arbitrary choice of the constant $\ p$ in the expression:%

\begin{equation}
J^{\ast k}=-\frac{\sqrt{2}c^{3}}{12G}\psi_{1k}^{0}|_{I}+p\frac{c^{3}}%
{G}\operatorname{Im}[\sigma\eth \overline{\sigma}+\frac{1}{2}\eth (\sigma
\overline{\sigma})]_{k}.\label{Jamb}%
\end{equation}
The default choices appear to be either 2, 1, or 0. The present work does not
influence or help resolve the ambiguity since to second order the expression
\begin{equation}
\operatorname{Im}[\sigma\eth \overline{\sigma}+\frac{1}{2}\eth (\sigma
\overline{\sigma})]_{k}=i\frac{3\cdot(24)^{2}\sqrt{2}}{10}(\xi^{il}%
\overline{\xi}^{ij}+\overline{\xi}^{il}\xi^{ij})\epsilon_{ljk},\label{zero}%
\end{equation}
vanishes.

One might have thought that our flux law, (still omitting the Maxwell field),%

\begin{align}
J^{k}{}^{\,\prime}  & =-\frac{(24)^{2}c^{3}}{5G}(\xi_{R}^{il}v_{R}^{ij}%
+\xi_{I}^{il}v_{I}^{ij})\epsilon_{ljk},\label{Jamb'}\\
J^{i}  & =-\frac{\sqrt{2}c^{3}}{12G}\psi_{1i}^{0}|_{I}\nonumber
\end{align}
would have an ambiguity in the flux, i.e., a total derivative, arising from
the use of the chain-rule, e.g.,
\[
\epsilon_{ljk}\xi_{R}^{il}v_{R}^{ij}=\epsilon_{ljk}(\xi_{R}^{il}\xi_{R}%
^{ij})^{\prime}-\epsilon_{ljk}v_{R}^{il}\xi_{R}^{ij}=-\epsilon_{ljk}v_{R}%
^{il}\xi_{R}^{ij}.
\]
This apparent ambiguity disappears since these total derivatives are
equivalent to the expression in Eq.(\ref{zero}) and again vanish identically .

It is possible that if our calculations were repeated, but done to third
order, the ambiguities could be resolved.

\end{document}